\documentclass[linenumbers,preprint]{aastex631}

\usepackage{comment}
\usepackage{hyperref}
\usepackage{amsmath}
\usepackage{cleveref}
\usepackage{enumitem}
\usepackage{graphicx}
\usepackage{lmodern}

\accepted{7 July 2023}

\submitjournal{AJ}

\shorttitle{Images and Spectroscopy of RW Cep}
\shortauthors{Anugu et al.}
\graphicspath{{./}{figures/}}

\begin{document}
\nolinenumbers

\title{The Great Dimming of the hypergiant star RW Cephei: \\
CHARA Array images and spectral analysis}

\correspondingauthor{Douglas R. Gies} \email{dgies@gsu.edu}

\author[0000-0002-2208-6541]{Narsireddy Anugu}
\affiliation{The CHARA Array of Georgia State University, Mount Wilson Observatory, Mount Wilson, CA 91023, USA}

\author[0000-0002-5074-1128]{Fabien Baron}
\affiliation{Center for High Angular Resolution Astronomy and Department 
of Physics and Astronomy, Georgia State University, P.O. Box 5060, Atlanta,
GA 30302-5060, USA}

\author[0000-0001-8537-3583]{Douglas R. Gies}
\affiliation{Center for High Angular Resolution Astronomy and Department 
of Physics and Astronomy, Georgia State University, P.O. Box 5060, Atlanta,
GA 30302-5060, USA}

\author[0000-0001-9745-5834]{Cyprien Lanthermann}
\affiliation{The CHARA Array of Georgia State University, Mount Wilson Observatory, Mount Wilson, CA 91023, USA}

\author[0000-0001-5415-9189]{Gail H. Schaefer}
\affiliation{The CHARA Array of Georgia State University, Mount Wilson Observatory, Mount Wilson, CA 91023, USA}

\author[0000-0003-2075-5227]{Katherine A. Shepard}
\affiliation{Center for High Angular Resolution Astronomy and 
Department of Physics and Astronomy, Georgia State University, 
P.O. Box 5060, Atlanta, GA 30302-5060, USA} 

\author[0000-0002-0114-7915]{Theo ten Brummelaar}
\affiliation{The CHARA Array of Georgia State University, Mount Wilson Observatory, Mount Wilson, CA 91023, USA}

\author[0000-0002-3380-3307]{John D. Monnier}
\affiliation{Department of Astronomy, University of Michigan, Ann Arbor, MI 48109, USA}

\author[0000-0001-6017-8773]{Stefan Kraus}
\affiliation{Astrophysics Group, Department of Physics and Astronomy, University of Exeter, Exeter, EX4 4QL, UK}

\author[0000-0002-0493-4674]{Jean-Baptiste Le Bouquin}
\affiliation{Institut de Planetologie et d'Astrophysique de Grenoble, Grenoble 38058, France}

\author[0000-0001-9764-2357]{Claire L. Davies}
\affiliation{Astrophysics Group, Department of Physics and Astronomy, University of Exeter, Exeter, EX4 4QL, UK}

\author[0000-0002-1575-4310]{Jacob Ennis}
\affiliation{Department of Astronomy, University of Michigan, Ann Arbor, MI 48109, USA}

\author[0000-0002-3003-3183]{Tyler Gardner}
\affiliation{Astrophysics Group, Department of Physics and Astronomy, University of Exeter, Exeter, EX4 4QL, UK}

\author[0000-0001-8837-7045]{Aaron Labdon}
\affiliation{European Southern Observatory, Casilla 19001, Santiago 19, Chile}

\author[0000-0002-9288-3482]{Rachael M. Roettenbacher}
\affiliation{Department of Astronomy, University of Michigan, Ann Arbor, MI 48109, USA}

\author[0000-0001-5980-0246]{Benjamin R. Setterholm}
\affiliation{Department of Climate and Space Sciences and Engineering, University of Michigan, Ann Arbor, MI 48109, USA}

\author{Wolfgang Vollmann}
\affiliation{Bundesdeutsche Arbeitsgemeinschaft Veraenderliche Sterne, Munsterdamm 90, D-12169 Berlin, Germany} 
\affiliation{American Association of Variable Star Observers, 185 Alewife Brook Parkway, \#410, Cambridge, MA 02138, USA}

\author{Costantino Sigismondi}
\affiliation{International Center for Relativistic Astrophysics, Universit\`{a} Pontificia Regina Apostolorum and ITIS Galileo Ferraris, via R. Grazioli Lante 15A, 00195, Rome, Italy}

\begin{abstract} 
\nolinenumbers
The cool hypergiant star RW Cephei is currently in a deep photometric minimum
that began several years ago.  This event bears a strong similarity to the Great Dimming 
of the red supergiant Betelgeuse that occurred in 2019--2020.   We present the first
resolved  images of RW~Cephei that we obtained with the CHARA Array interferometer.
The angular diameter and Gaia distance estimates indicate a stellar radius of  
$900 - 1760 R_\odot$ which makes RW~Cephei one of the largest 
stars known in the Milky Way.  The reconstructed, near-infrared images show a striking 
asymmetry in the disk illumination with a bright patch offset from center and a darker zone to the west. 
The imaging results depend on assumptions made about the extended flux, and we 
present two cases with and without allowing extended emission.
We also present a recent near-infrared spectrum of RW~Cephei that  
demonstrates that the fading is much larger at visual wavelengths compared
to that at near-infrared wavelengths as expected for extinction by dust. 
We suggest that the star's dimming is the result of a recent surface mass ejection event
that created a dust cloud that now partially blocks the stellar photosphere. 
\end{abstract}


\null
\vspace{2.5 cm}

\section{Introduction} \label{sec:intro}

The recent Great Dimming of Betelgeuse provided an opportunity to 
study the dynamics of mass loss in a relatively nearby red supergiant 
(summarized by \citealt{Dupree2022}). During the months before the 
fading (2019 January to November), the spectrum of Betelgeuse indicated 
an outflow from the photosphere that was possibly related to a large 
convective upwelling \citep{Kravchenko2021, Jadlovsky2023}. This probably led to a 
surface mass ejection of a large gas cloud that cooled and formed  
dust and increased the visible band extinction \citep{Taniguchi2022}. 
\citet{Montarges2021} obtained angularly resolved images of Betelgeuse 
with VLT SPHERE-ZIMPOL around the deep minimum when the star had faded 
by 1.2 mag in the $V$-band (2019 December to 2020 March).  Their images 
showed that the southern hemisphere was much darker than in pre-minimum images 
suggesting that the fading was the result of partial extinction by a 
foreground dust cloud seen against a slightly cooler photospheric disk. 
The mass lost during this ejection was comparable to that for a full 
year of steady outflow \citep{Montarges2021} indicating that episodic 
mass ejections in supergiants constitute a significant fraction of
their total mass loss \citep{Humphreys2022, Massey2023}. 

The cool hypergiant RW~Cephei (HD~212466) is now presenting us with 
a second opportunity to explore episodic mass loss at high angular 
resolution. Its spectrum indicates a very high luminosity
(classified as  K2~0-Ia by \citealt{Keenan1989}), and the spectral line shapes 
suggest complex photospheric motions and outflow \citep{Merrill1956,Josselin2007}.
The star is a yellow semiregular variable, and it displays modest 
photometric variations on a timescale of about a year \citep{Percy2000}.  
However, recent photometric measurements by \citet{Vollmann2022}, 
\href{https://www.aavso.org/LCGv2/static.htm?DateFormat=Julian&RequestedBands=V&Grid=true&view=api.delim&ident=rw cep&fromjd=2457300&obscode=VOL&delimiter=@@@}{AAVSO observers}\footnote{https://www.aavso.org/LCGv2/}, 
and the Kamogata/Kiso/Kyoto Wide-field Survey\footnote{http://kws.cetus-net.org/$^\sim$maehara/VSdata.py} 
(KWS\; \citealt{Maehara2014}) 
show that RW~Cep is now undergoing its own great dimming episode (Figure 1).
By the end of 2022, RW~Cep had faded by 1.1 mag in $V$-band 
to become fainter than at any time in the last century. 
Furthermore, the star became redder (larger $V-I_c$) as it faded. 
At the time of writing (2023 June), it appears to have passed its point of minimum light 
and is slowly brightening again.

Visible-band spectra made in 2022 December by 
R.\ Leadbeater\footnote{https://www.cloudynights.com/topic/854288-rw-cephei-great-dimming/}
show a good match to that of a K4~I spectral template with an interstellar reddening of $E(B-V)=0.65$ mag.
High resolution spectra made by 
\href{http://www.threehillsobservatory.co.uk/astro/RW_Cep/rwcep_elodie_archive_THO_2022-12-19_Halpha.png}{R.\ Leadbeater}\footnote{http://www.threehillsobservatory.co.uk/astro/RW\_Cep/rwcep\_elodie\_archive\_THO\_2022-12-19\_Halpha.png} 
and by 
\href{http://www.spectro-aras.com/forum/viewtopic.php?f=42&t=3057$#$p17405}{J. Guarro Fl\'{o}}\footnote{http://www.spectro-aras.com/forum/viewtopic.php?f=42\&t=3057$\#$p17405}
show evidence of a narrow H$\alpha$ emission line that was absent in ELODIE spectra of the 
star that were made between 1999 and 2005. This emission may be associated with excess mass loss during
the current episode.  The star has a strong infrared flux excess that forms in a dust envelope 
\citep{Gehrz1971, RowanRobinson1982, Jones2023}. It appears slightly extended (diameter $\approx 1$ arcsec) 
in  high resolution, mid-infrared images 
(see Fig.\ A.3 in \citealt{Shenoy2016} and Fig.\ 1 in \citealt{Jones2023})
indicating a long history of mass loss. 
 
\begin{figure*}[h!]
\gridline{\fig{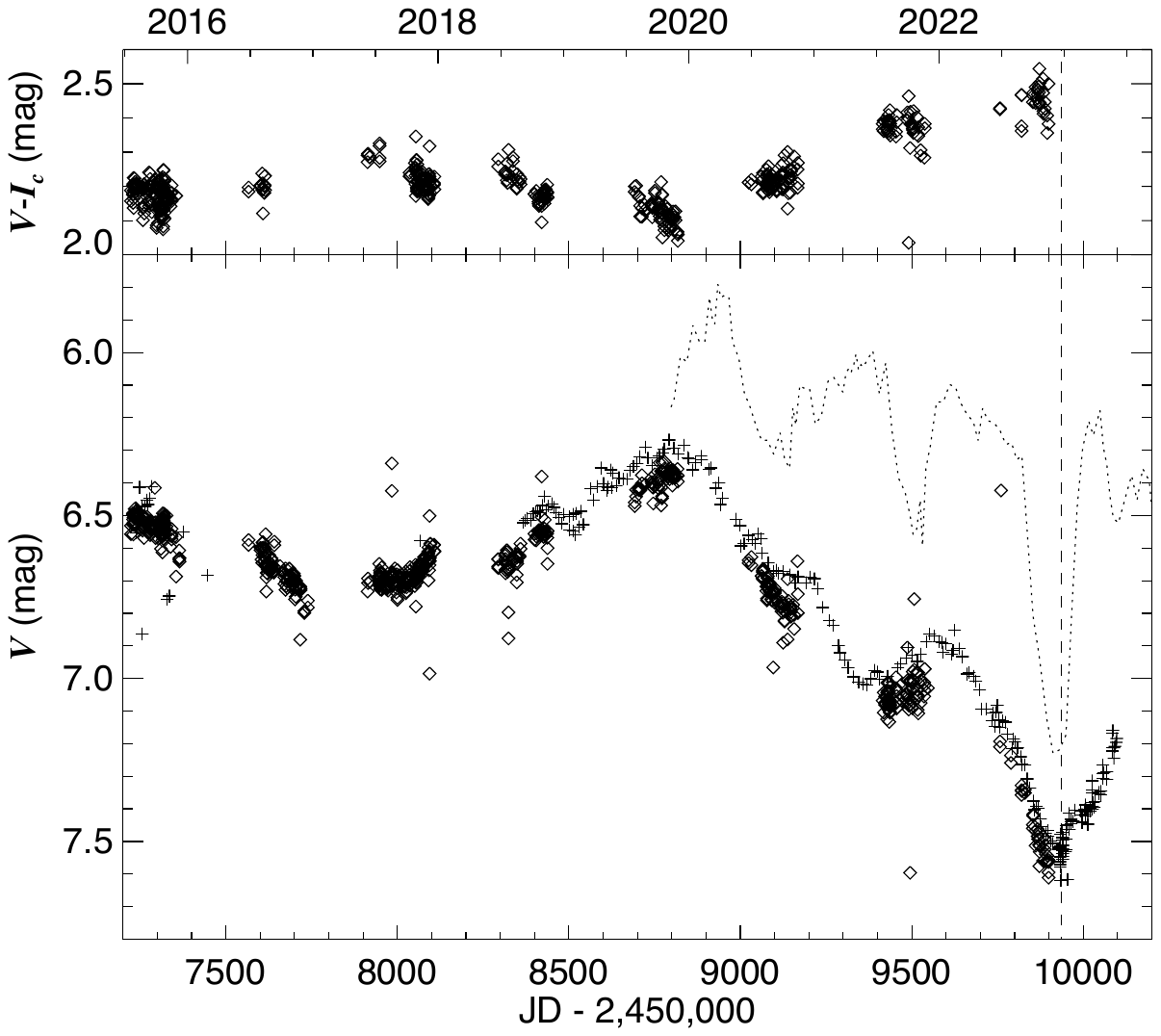}{1.0\textwidth}{} }
\caption{The $V$-band light curve (lower panel) and $V-I_c$ color index (upper panel) 
 of RW~Cep over the past seven years. The lower panel shows the 
 $V$-band light curve from the AAVSO archive (plus signs; 
 mainly from observations made by W.\ Vollman) and the KWS (diamond symbols). 
 The KWS includes estimates of the Johnson-Cousins $I_c$-band magnitude, 
 and the top panel shows the time evolution of the $V-I_c$ color index. 
 The top axis gives the calendar date (BY) of observation.
 The vertical dashed line indicates the date of the CHARA Array observation.
 For comparison purposes, the dotted line in the lower panel shows the Great Dimming 
 of Betelgeuse from \citet{Taniguchi2022} for their 0.51 $\mu$m band photometry 
 (7 point averages, offset by $+1040$ days in time and by $+5.2$ in magnitude).}
\label{fig1}
\end{figure*}

\placefigure{fig1}

Here we present the first interferometric images of RW~Cep that we
made recently with the CHARA Array, and these show striking similarities 
to the asymmetries seen in the VLT images of Betelgeuse.  We describe 
the interferometric observations and derived images in Section 2, and 
we present a recently obtained near-infrared flux spectrum in Section 3. 
A comparison is made of the spectral energy distribution before and 
during the fading event in Section 4.  We discuss the implications of 
these observations for models of dimming and mass ejection in Section 5. 



\section{Interferometric Images} \label{sec:images}

We obtained a single observation of RW~Cep with the 
Center for High Angular Resolution Astronomy (CHARA) Array 
\citep{tenBrummelaar2005, Schaefer2020}  on 2022 December 23 UT.  
We used only five of the six Array telescopes because the S1 telescope 
had insufficient available delay length for the star's position in the 
north-western sky at the time of the observations.
We used the dual beam combiners
MIRC-X \citep{Anugu2020} for the near-infrared $H$-band (1.50 to 1.74 $\mu$m)
and MYSTIC for the $K$-band (2.00 to 2.37 $\mu$m)
\citep{Monnier2018, Setterholm2022}.  
The observations were made with
a spectral resolving power of $R=190$ and 100 for MIRC-X and 
MYSTIC, respectively.  The nominal angular resolution is 
approximately 0.5 and 0.6 milliarcsec (mas), respectively. 
These beam combiners use the telescopes of the Array to 
collect interferometric fringe measurements for a large
range in baseline over much of the $(u,v)$ spatial frequency plane (Figure 2).

\begin{figure*}[h]
\plottwo{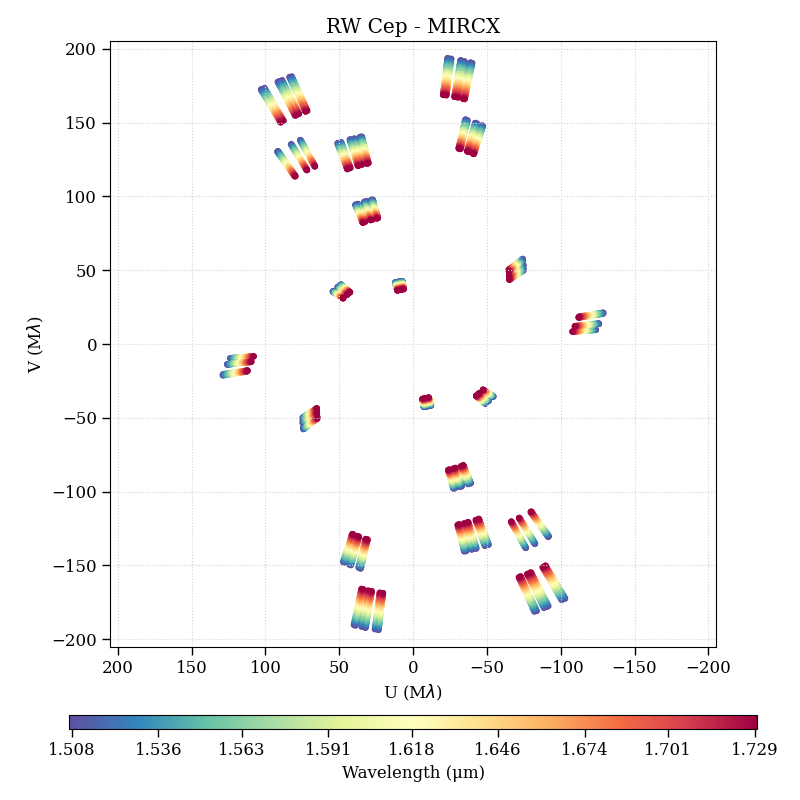}{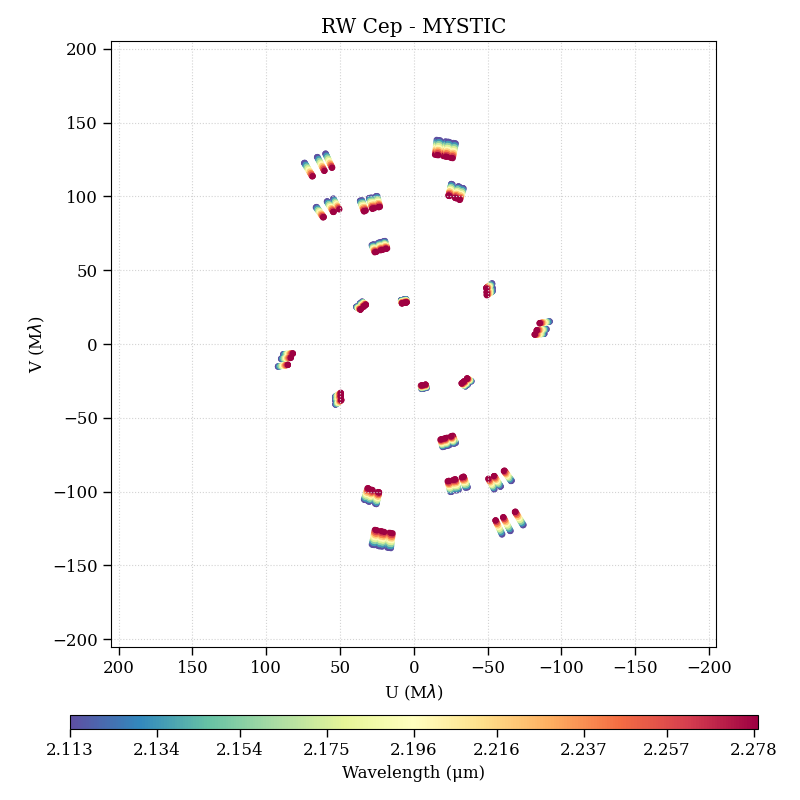}
\caption{The spatial frequency coverage in the $(u,v)$ plane for the 
$H$-band (left) and $K$-band (right) observations. The lower color 
legends show the corresponding wavelength channels.}
\label{fig2}
\end{figure*}

\placefigure{fig2}

The measurements were reduced using the standard MIRC-X/MYSTIC 
pipeline (version 1.4.0; \citealt{Anugu2020}).  Calibrator 
observations were made of HD~219080 (before the science target) and 
were used to correct for atmospheric and instrumental effects to obtain 
absolute-calibrated visibilities $V^2$, closure phases (CP), and triple amplitudes (T3A).  
The calibrator diameter (0.69 mas for a uniform disk) was 
adopted from the JMMC Stellar Diameters Catalog (JSDC;  
\citealt{Bourges2017}).  The derived visibilities and closure 
phases are shown in Figures 3 and 4 for the $H$- and $K$-bands, respectively.
The star is clearly resolved in both bands (visibility declining with larger spatial frequency), 
and the data show evidence of an asymmetric flux distribution (non-zero and non-$\pi$ closure phase).

\begin{figure*}[ht!]
\plottwo{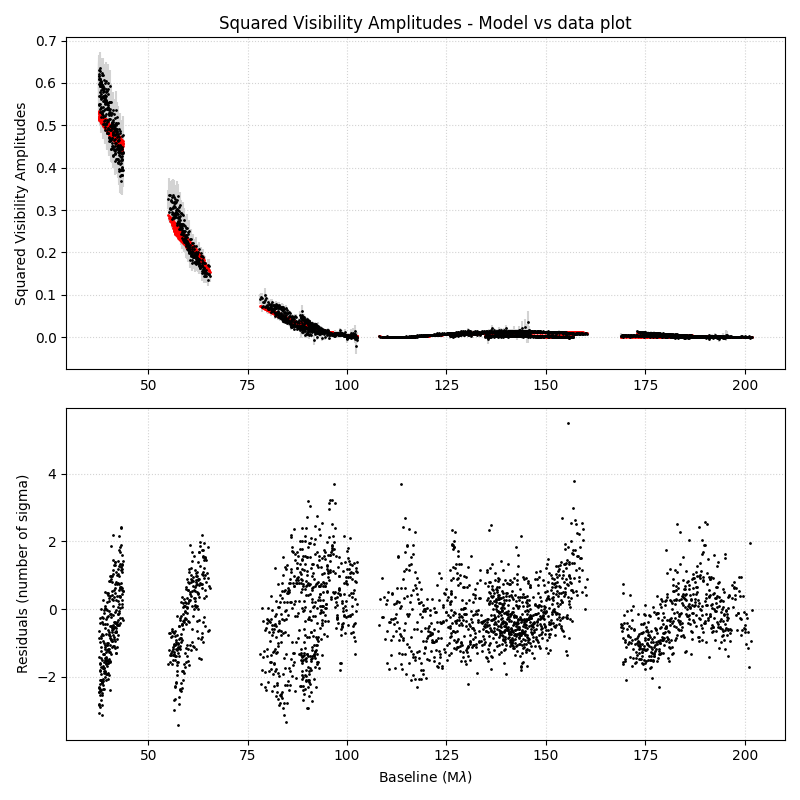}{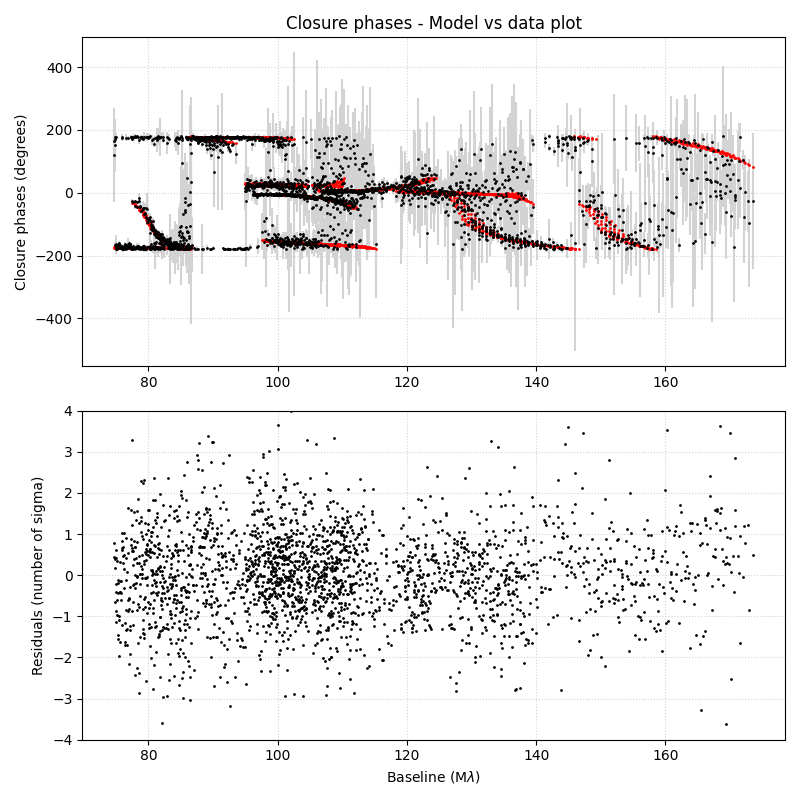}
\caption{The $H$-band visibilities (left) and closure phases (right)
for the telescope pair and triplet MIRC-X observations of RW~Cep.
The black symbols with vertical error bars are the measurements
and the red dots represent the corresponding fits from the SQUEEZE reconstructed images. 
The lower panels show the residuals from the SQUEEZE fits. The fits from the OITOOLS 
and SURFING reconstructions yield residuals that are qualitatively similar in appearance.}
\label{fig3}
\end{figure*}

\placefigure{fig3}

\begin{figure*}[hb!]
\plottwo{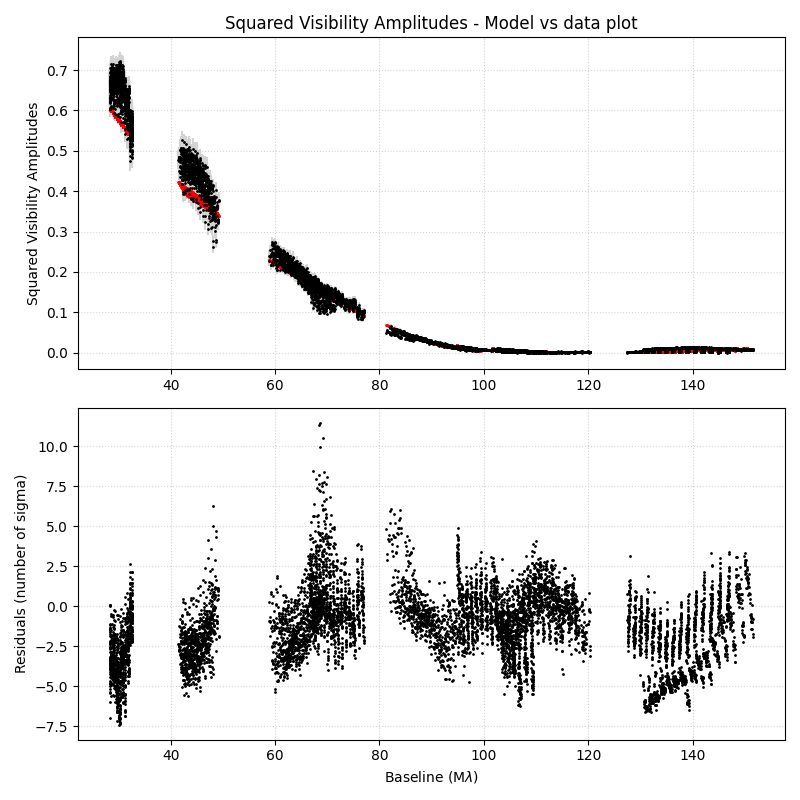}{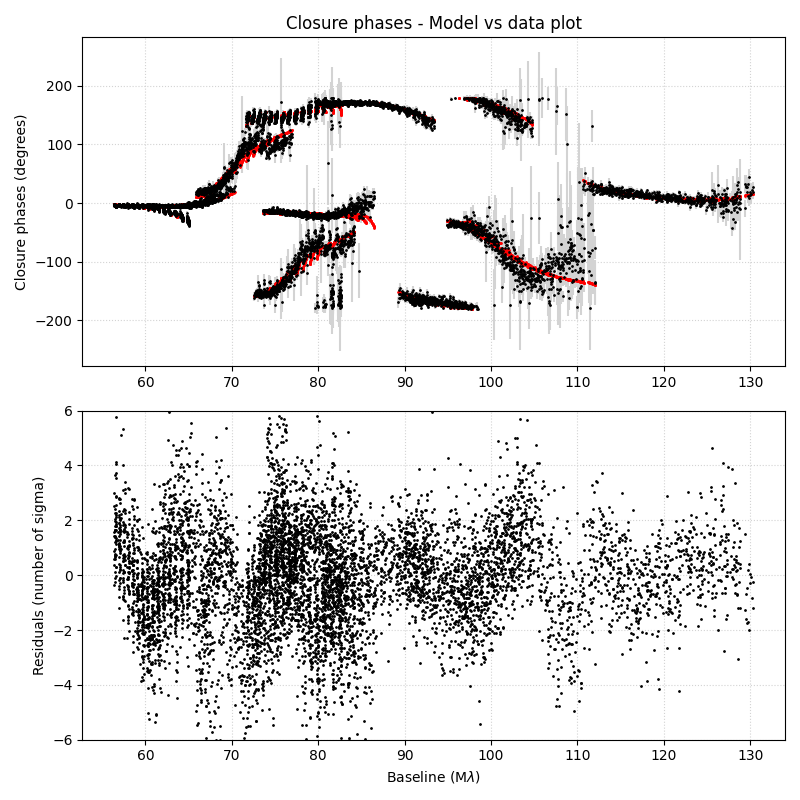}
\caption{The $K$-band visibilities (left) and closure phases (right)
in the same format as Figure 3.}
\label{fig4}
\end{figure*}

\placefigure{fig4}

We first fit the interferometric visibilities $V^2$ using an analytical model 
of a uniformly bright circular disk for the star with an incoherent background flux 
on larger spatial scales over-resolved in the CHARA Array observations. 
The fits were made over different wavelength ranges using the code 
PMOIRED\footnote{https://github.com/amerand/PMOIRED}
\citep{Merand2022}. Table~1 lists the derived values of the background 
flux fraction, uniform disk angular diameter $\theta = \theta _{UD}$, and reduced 
chi-squared $\chi ^2 _\nu$ of the fit of the visibilities.  The first row gives the 
$H$-band fits of the MIRC-X measurements, and the second and third rows give the 
$K$-band results from MYSTIC in the $K$-band pseudo-continuum and in the CO bands, 
respectively (see Figure 8 below). 
The angular diameter is found to increase in size with wavelength and 
appears to be significantly larger in the CO bands (row 3).  This is evidence
of an extended atmospheric extension in the CO bands that is also 
observed in other cool, luminous stars \citep{Perrin2005, Tsuji2006, LeBouquin2009, Ohnaka2011, Montarges2014}.
There is an extended background flux that generally forms 
a larger fraction of the total flux at longer wavelength. We suspect that 
this flux originates in extended dust emission that reaches an angular size of order 
1 arcsec at 10 $\mu$m \citep{Shenoy2016, Jones2023}. 

\begin{deluxetable*}{ccccccc}[h]
\tablecaption{Angular Diameter Estimates \label{tab1} }
\tablewidth{0pt}
\tablehead{
\colhead{Method} & 
\colhead{Wavelength} & 
\colhead{Background}  &
\colhead{$\theta$} &
\colhead{$\chi ^2 _\nu$} &
\colhead{$\chi ^2 _\nu$} &
\colhead{$\chi ^2 _\nu$}
\\ 
\colhead{}  & 
\colhead{($\mu$m)}  & 
\colhead{Fraction}  &   
\colhead{(mas)}  & 
\colhead{($V^2$)}  &
\colhead{(CP)} &
\colhead{(T3A)} 
}
\startdata
PMOIRED UD & 1.50 -- 1.72 & 0.09 $\pm$ 0.02 & 2.44 $\pm$ 0.02 & 3.8 & \nodata & \nodata \\
PMOIRED UD &1.98 -- 2.29 & 0.17 $\pm$ 0.01 & 2.63 $\pm$ 0.02 & 4.2  & \nodata & \nodata \\
PMOIRED UD &2.31 -- 2.37 & 0.15 $\pm$ 0.01 & 3.21 $\pm$ 0.02 & 2.6  & \nodata & \nodata \\
\hline
SQUEEZE & 1.50 -- 1.72 & 0.19 & 2.26 -- 2.69   & 1.37  & 1.24  & 1.35 \\
SQUEEZE &1.98 -- 2.29 & 0.17 & 2.08 -- 2.66    & 1.12  & 1.73  & 1.01 \\
OITOOLS & 1.50 -- 1.72 & 0.09 & 2.41   & 1.29  & 1.10 & 0.54 \\
OITOOLS &1.98 -- 2.29 & 0.08 & 2.35   &  1.26  & 2.28 & 0.59 \\
SURFING & 1.50 -- 1.72 &   0.08  & 2.45 & 2.77 & 2.92 &1.13 \\
SURFING & 2.11 -- 2.28 &   0.06 & 2.44 & 1.32 & 7.91 & 0.88 \\
\hline
SED fit & 0.35 -- 2.20 & 0 & $2.58 \pm 0.16$ & \nodata & \nodata & \nodata \\ 
\enddata
\end{deluxetable*}

The relatively high quality of the visibility and closure phase measurements 
encouraged us to derive aperture synthesis images that make good fits of 
the observations.  We caution at the outset that this 
data set is not ideal for image reconstruction.  The observations were made only
over a duration of one hour when the star was already at a large hour angle,
and only five of the six telescopes were available.  Consequently, the 
$(u,v)$ spatial frequency coverage is under-represented in some sky 
orientations, and the effective angular resolution is better in the north - 
south directions compared to the east - west directions (Figure 2). 
Consequently, the spatial resolution in the reconstructed images varies 
with position angle and is poor at position angles near 
$+60^\circ$ and $-40^\circ$ (both $\pm 180^\circ$).  
Furthermore, the star has a relatively small angular size, and any 
small-scale structured flux in the extended emission will complicate 
the image reconstruction of the star itself.

We first used the SQUEEZE image reconstruction 
software\footnote{https://github.com/fabienbaron/squeeze} 
\citep{Baron2010,Baron2012} to make the images.
Positivity was enforced, and we used the edge-preserving $\ell_2-\ell_1$ regularization 
\citep{LeBesnerais2008} with the hyperparameter weight determined by the classical $L$-curve method. 
To avoid getting trapped in local minima in the solution space, 50 initial reconstructions 
were started from random images.  The solutions converged into images with similar appearance, 
and these trials were then co-registered and averaged to obtain representative images.
The reduced $\chi^2$ values were 1.37 for the MIRC-X data fits and 3.7 for the MYSTIC data fits, 
suggesting much stronger chromaticity (wavelength dependence) for the latter.
For the MYSTIC data, we removed the spectral channels containing the CO bands 
to perform an image reconstruction in the $K$-band pseudo-continuum. 
We obtained a lower reduced $\chi^2 \sim 1.12$ by omitting  
those long wavelength bins that record the CO bands.

The SQUEEZE reconstructed images are shown in the top panels of Figure~5 for each of 
the $H$-band and wavelength restricted $K$-band observations.  These are 
$6.4 \times 6.4$ mas images ($64 \times 64$ pixels) with an orientation of north to 
the top and east to the left. There is a clear asymmetry evident
in the images with a brighter zone towards the north-east limb
and a darker (and possibly extended) zone towards the western side.
We experimented with several other choices of regularizer for the image reconstruction,
and the same large scale asymmetry appears in those images. 
The mean background and diameter estimates from the SQUEEZE images are given in Table~1.

The non-spherical shape and possible limb extensions that characterize the SQUEEZE 
images may have a physical rather than instrumental origin.  Models of mass loss in 
red supergiant stars by \citet{Hofner2019} indicate that such stars may have extended 
clumpy regions that create shell-like features, and observations of the hypergiant 
VY~CMa by \citet{Humphreys2021} show the presence of clumps and knot structures close
to the star. \citet{LopezAriste2023} obtained spectropolarimetry of the hypergiant 
star $\mu$~Cep that they interpret in terms of rising convective plumes that reach a
radius of $1.1 R_\star$. Thus, the irregular shape of RW~Cep in the SQUEEZE image 
reconstructions may be due to the combined effects of extended plume emission and 
localized dust emission and absorption (together creating only a modest change in overall flux;
see Table 2 below).  

We were concerned that the boxy image structure might be 
due to the limited $(u,v)$ coverage of the observations (see Figure~2), 
so we performed a numerical test to check if the non-spherical appearance is due to the star itself.
We created a model of a spherical, limb-darkened disk (power law) using the angular diameter
$\theta$ determined from the OITOOLS reconstructions described below (see Table~1). Then we used 
these model images to generate the OIFITS data sets that would have been observed for these simple
disks. We performed SQUEEZE reconstructions from the model data using the same $(u,v)$ coverage 
and noise levels associated with the observations. The resulting SQUEEZE images are shown
in the bottom row of Figure~5, and these appear more or less circular as expected.  
These tests indicate that the unusual shape of RW~Cep in the SQUEEZE images reconstructed 
from the observations probably does not have an instrumental explanation, but that the stellar shape 
is sculpted by dynamical processes in its outer layers. 

It is worthwhile considering how the star would appear if the image reconstruction
instead is confined to within the stellar radius, and the star is surrounded by a diffuse, 
over-resolved background light. 
We did this by making sets of images that constrain the structured flux to 
fall within a circle defined by the stellar photosphere.  We adjusted the uncertainties in 
the measurements in these cases by adding a $10\%$ relative error and a 0.0002 additive
correction for $\triangle V^2$ and adding a minimum error of 1 degree for the closure phase errors. 
These revisions account for possible systematic uncertainties.   
A set of OITOOLS images were obtained using the OITOOLS.jl software suite\footnote{https://github.com/fabienbaron/OITOOLS.jl}. 
The initial starting images consisted of the best-fitting uniform disks derived for the MIRC-X and MYSTIC datasets. 
The regularization was set up to use a combination of image centering, compactness and 
$\ell_1$-$\ell_2$ edge-preserving smoothness \citep{Thiebaut2017}, 
where the compactness prior was set as the starting image.  The $H$ and $K$-band images 
from the OITOOLS reconstructions (for $128 \times 128$ pixels) are shown in 
the top panels of Figure~6, and the associated background and angular diameter 
estimates are given in Table~1.  The star appears to be larger and more circular using 
this method, but some of the same flux asymmetries found in the SQUEEZE images 
are also recovered here but with lower contrast. 

\begin{figure}[h!]
\vspace{-3.0 cm}
\begin{minipage}[h]{0.47\linewidth}
\begin{center}
\includegraphics[width=0.9\linewidth]{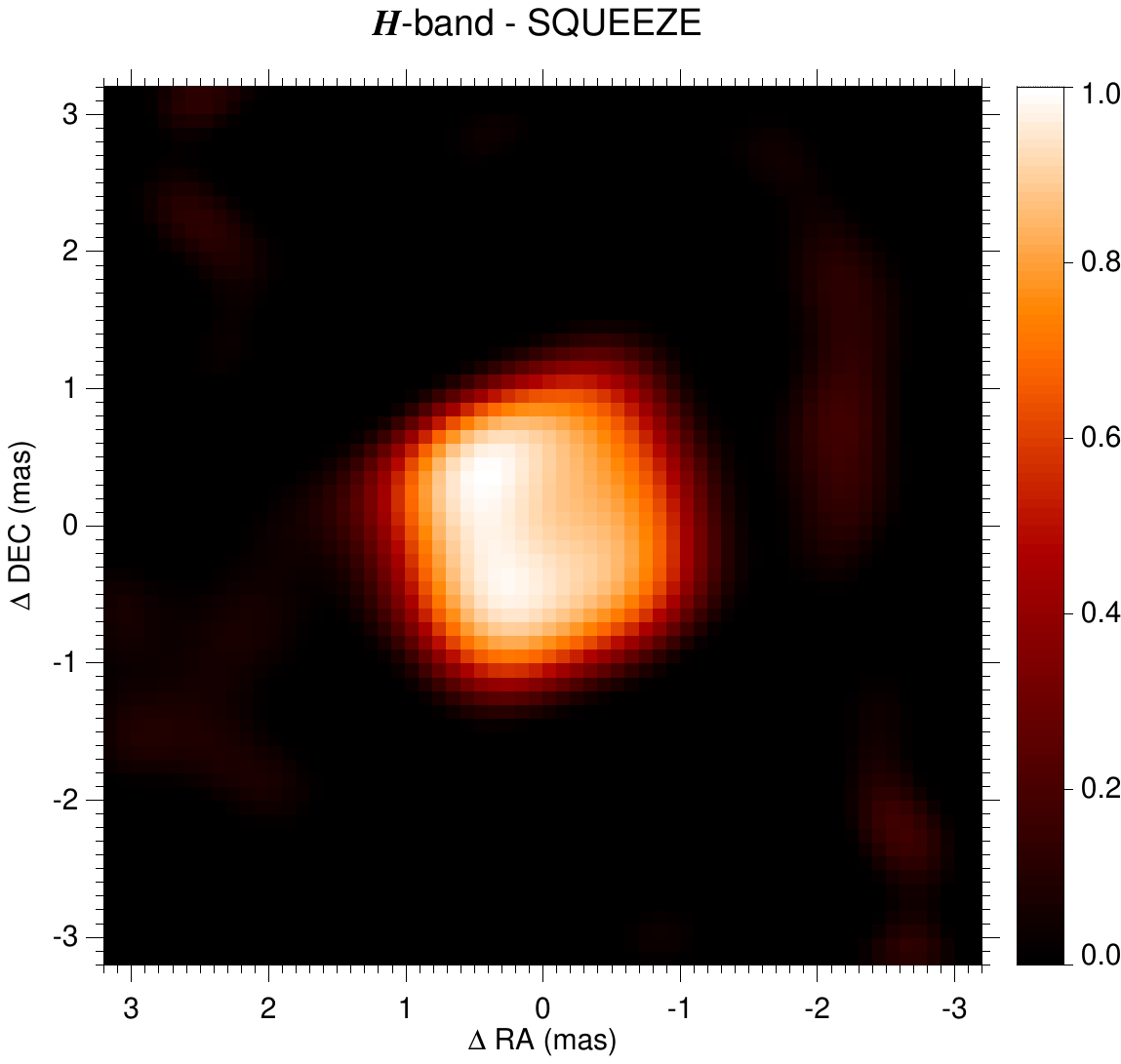}
\end{center} 
\end{minipage}
\hfill
\vspace{0.2 cm}
\begin{minipage}[h]{0.47\linewidth}
\begin{center}
\includegraphics[width=0.9\linewidth]{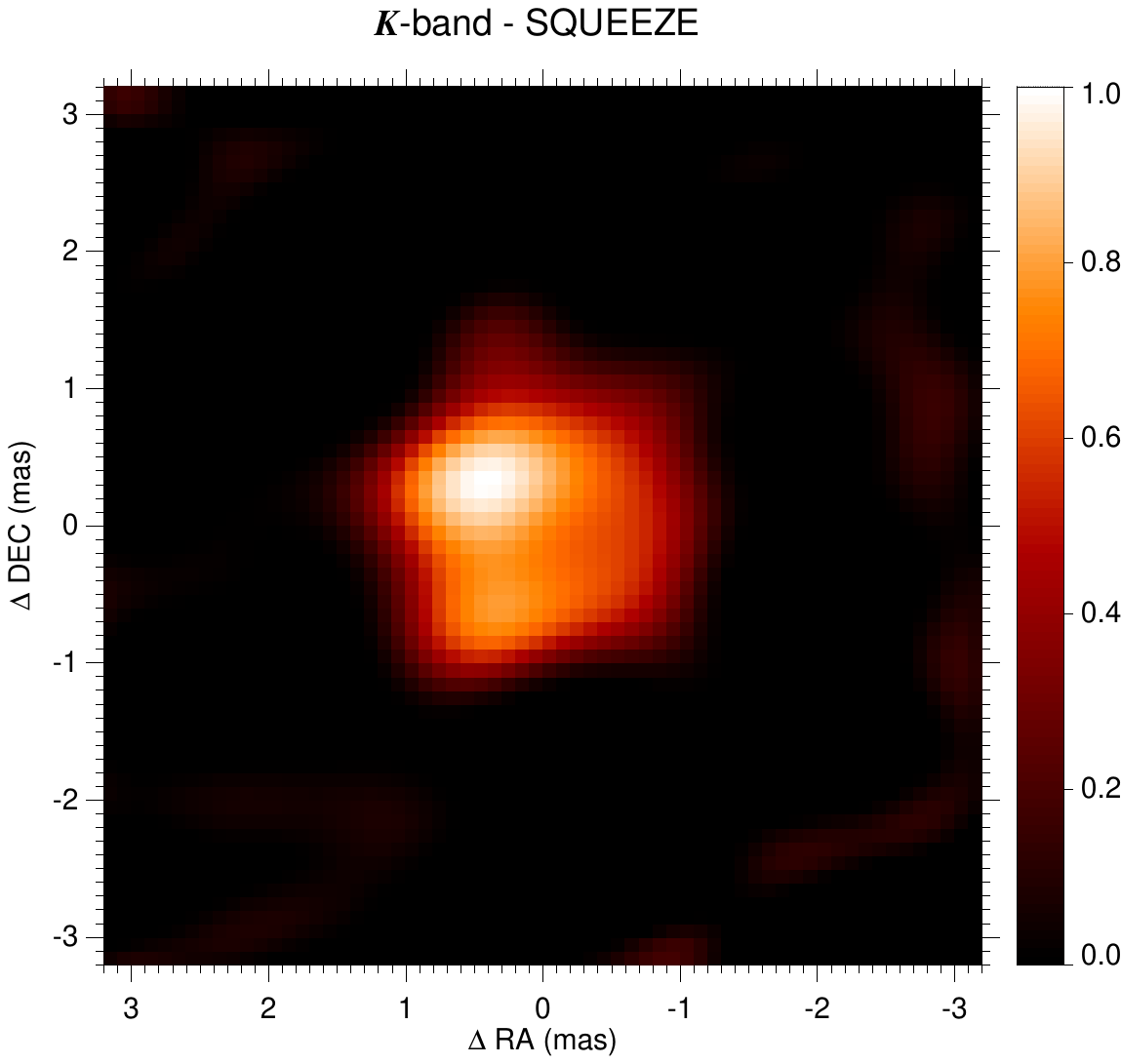}
\end{center}
\end{minipage}
\vfill
\vspace{-2.5 cm}
\begin{minipage}[h]{0.47\linewidth}
\begin{center}
\includegraphics[width=0.9\linewidth]{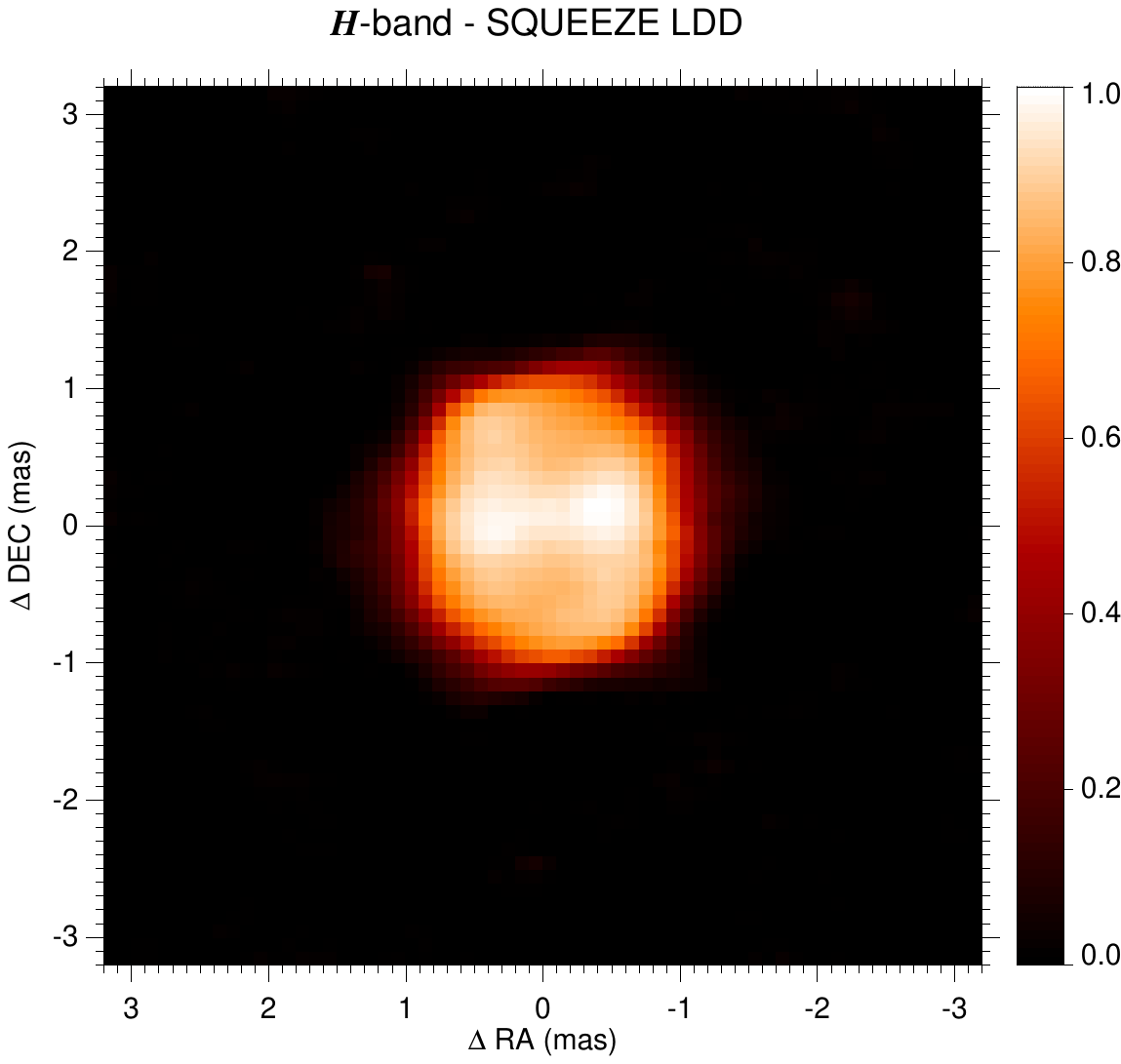}
\end{center} 
\end{minipage}
\hfill
\vspace{0.2 cm}
\begin{minipage}[h]{0.47\linewidth}
\begin{center}
\includegraphics[width=0.9\linewidth]{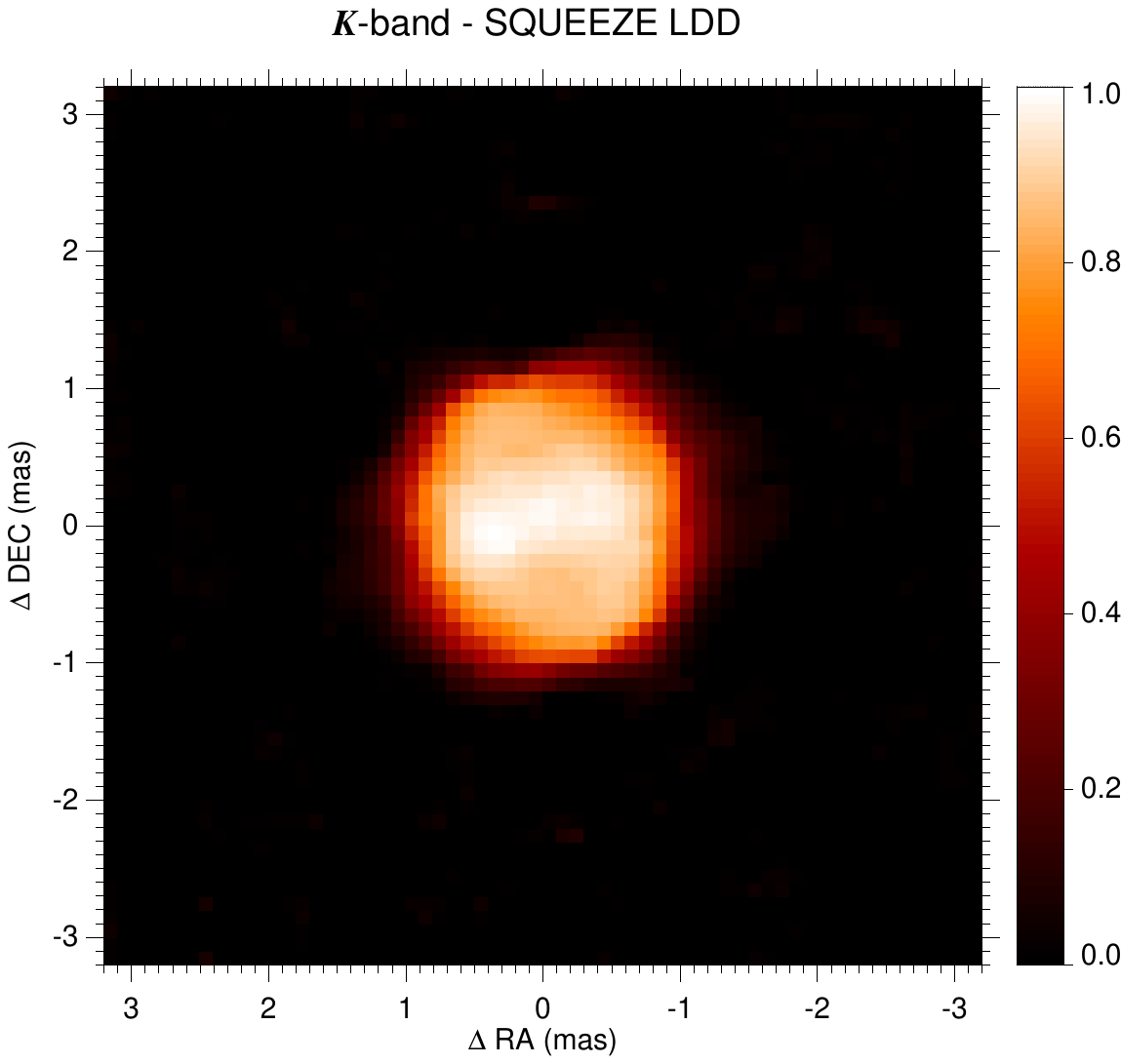}
\end{center}
\end{minipage}
\caption{The $H$-band (top left) and $K$-band (top right) images 
of RW~Cep made using the SQUEEZE algorithm. 
The color bar at right shows the correspondence between 
broadband specific intensity (normalized to the brightest pixel) and image color. 
The lower panels show the corresponding images reconstructed from the same 
$(u,v)$ sampling of a model image of a limb darkened star.
}
\label{fig5}
\end{figure}

\begin{figure}[h!]
\vspace{-3.0 cm}
\begin{minipage}[h]{0.47\linewidth}
\begin{center}
\includegraphics[width=0.9\linewidth]{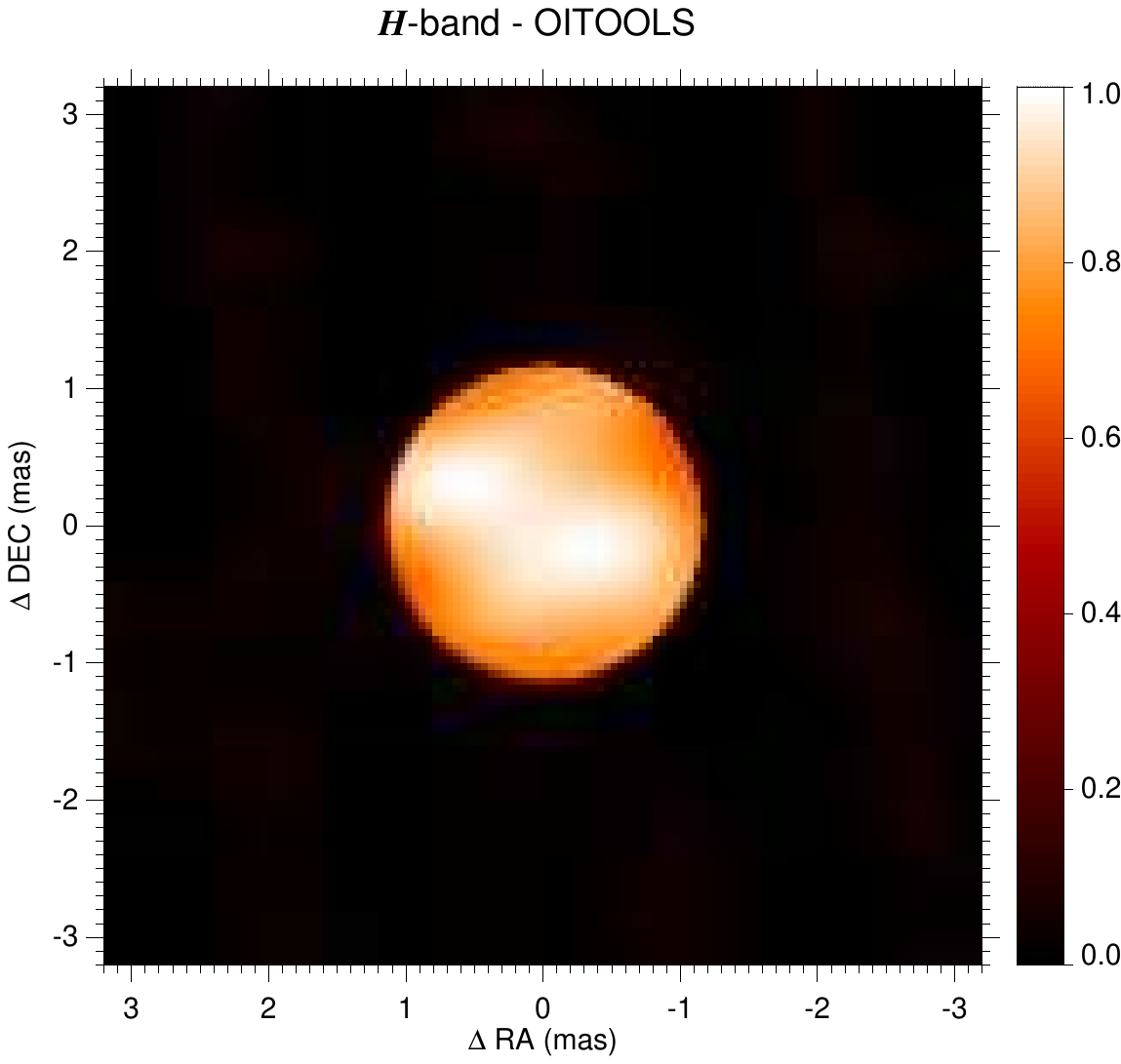}
\end{center} 
\end{minipage}
\hfill
\vspace{0.2 cm}
\begin{minipage}[h]{0.47\linewidth}
\begin{center}
\includegraphics[width=0.9\linewidth]{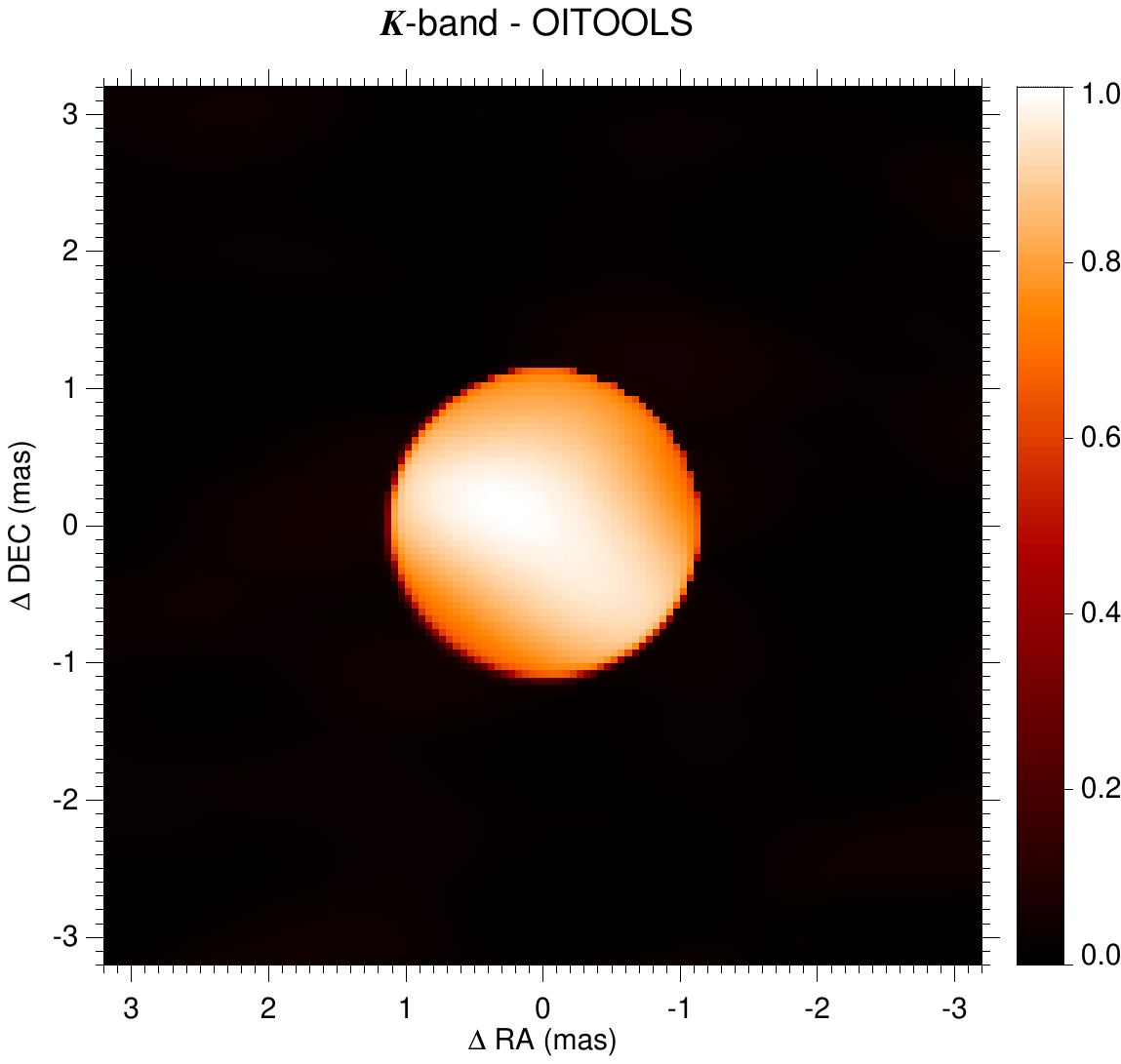}
\end{center}
\end{minipage}
\vfill
\vspace{-2.5 cm}
\begin{minipage}[h]{0.47\linewidth}
\begin{center}
\includegraphics[width=0.9\linewidth]{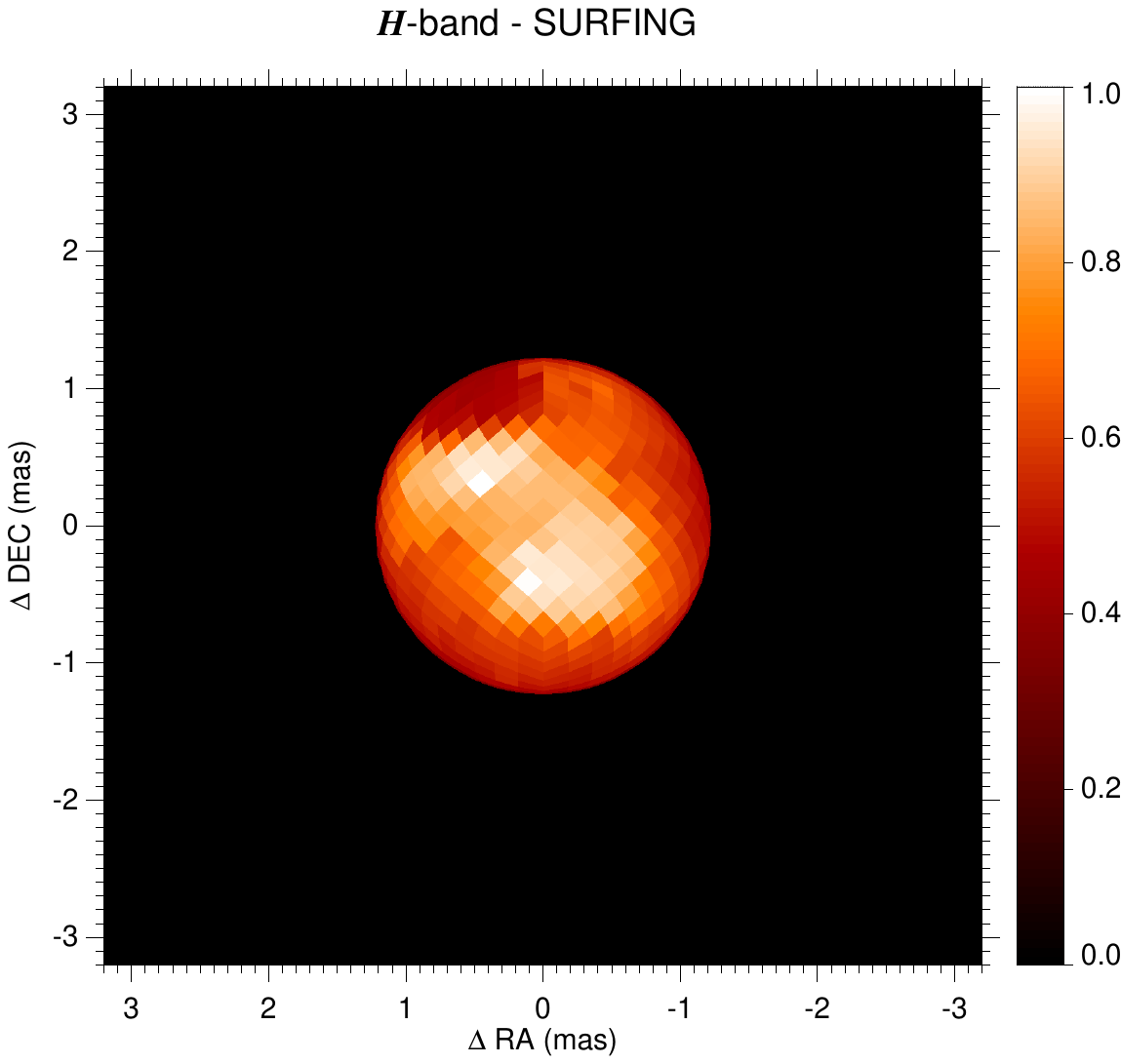}
\end{center}
\end{minipage}
\hfill
\begin{minipage}[h]{0.47\linewidth}
\begin{center}
\includegraphics[width=0.9\linewidth]{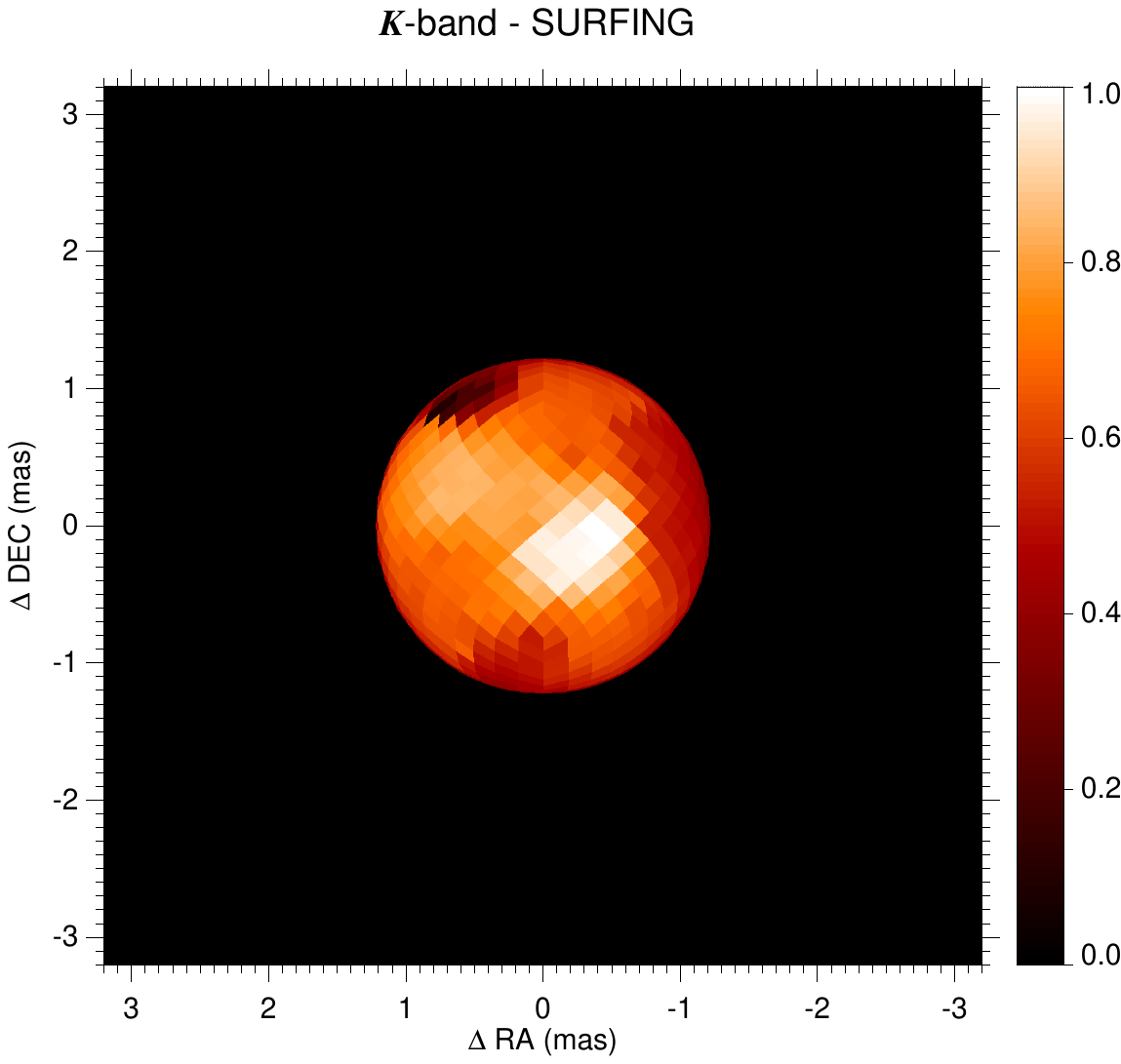}
\end{center}
\end{minipage}
\caption{The $H$-band (left) and $K$-band (right) images 
of RW~Cep made using the OITOOLS (top), and SURFING (bottom) algorithms
based upon fitting the specific intensity within the stellar disk only.
The color bar at right shows the correspondence between 
broadband specific intensity (normalized to the brightest pixel) and image color. }
\label{fig6}
\end{figure}


One more set of image reconstructions
were made using the SURFING algorithm \citep{Roettenbacher2016, Martinez2021} that 
assigns a specific intensity to each element on the three-dimensional surface of the star.  The first step 
was to find a best-fit limb-darkened angular diameter that acts as the outer boundary on 
the assigned flux. The best fit diameters $\theta = \theta_{LD}$ are listed in Table 1, and 
these were derived assuming a power law limb-darkening relation, $I(\mu)/I(\mu =1) = \mu ^ \alpha$, 
where $\mu$ is the cosine of the angle between surface normal and line of sight and 
$\alpha = 0.26$.  The SURFING algorithm was then applied iteratively to solve for the 
surface element brightness (in this case only for elements on the visible hemisphere). 
We show in the lower panels of Figure~6 one pair of images among the final set of 
walker solutions ($1024 \times 1024$ pixels of size 0.005 mas, inset into a uniform 
background zone).  The $H$ and $K$-band images appear similar to each other and 
show a bright patch offset from center and a darkening to the western limb.   
We also created a set of images using the ROTIR code \citep{Martinez2021} 
that likewise assigns flux to surface patches on a rotating star, and these 
images are qualitatively similar to those derived using SURFING. 

The SQUEEZE, OITOOLS, and SURFING images show some similarities but also some 
significant differences in appearance. The SQUEEZE images (Fig.~5) were made with the fewest 
assumptions about the expected appearance.  The star in the SQUEEZE images is non-circular
and boxy in appearance with sides tilted by about $20^\circ$ relative to north. 
The star appears darker on its western limb, and the brightest zone is positioned 
north-east of center.  The $K$-band image shows greater contrast across the disk 
and the limb is spread over a larger span in radius. We show in Section 3 below  
that dust emission begins to become a flux contributor in the $K$-band, and the dust opacity can 
create both emission (off of the stellar disk) and absorption (projected against 
the disk). 

The OITOOLS and SURFING images (Fig.~6) restrain the reconstructed flux to the star. 
They show darker limbs (especially the western limb) coincident with the boxy sides 
seen in the SQUEEZE images.  There is also a bright off-center patch that appears
towards the north-east (south-west) in the OITOOLS (SURFING) $K$-band images, 
while two offset patches appear in the $H$-band images.  The differences in 
bright zone position may result from the neglect of structured off-disk 
light in these two algorithms.  Note that the total amount of off-disk flux is
about two times larger in the SQUEEZE reconstructed images compared to the 
OITOOLS and SURFING images (see the background fraction given in Table~1). 

The surface intensity distribution of red supergiants is probably dominated by hot, rising 
convection cells \citep{Chiavassa2009, Norris2021}, but in the case of hypergiants, 
mass loss becomes the dominant process that shapes the intensity distribution
\citep{Hofner2019, Humphreys2022}.  We expect that the local mass-loss rate 
may vary with position on the star due to the kinematics and radiation of hot
convective cells, and the observational consequences may be especially 
important at the stellar limb where hotter gas can create spatially extended 
emission.  The SQUEEZE images were made without any geometric assumptions 
about spherical symmetry, and the irregular stellar shape in these images may reflect 
the spatial variation in mass-loss rate.   


\vspace{2 cm}

\section{Near-Infrared Spectroscopy} \label{sec:spectrum}

We obtained complementary near-infrared (NIR) spectroscopy of RW~Cep
using the TripleSpec instrument at the 3.5~m telescope of the Apache Point 
Observatory \citep{Wilson2004}.  TripleSpec records the NIR spectrum over 
the wavelength range of 0.9 to 2.5 $\mu$m with a spectral resolving power of $R=3500$. 
The observations were obtained on 2023 January 9 and 12 in good sky conditions. 
We made sets of the standard ABBA exposure nodding pattern for slit offset 
positions A and B for subtraction of the sky background.  In order to avoid saturation 
of the detector for the bright flux of RW~Cep, the telescope was defocussed 
to create a broad (and double-peaked) spatial profile across the spectrograph 
slit.  We made multiple observations of RW~Cep and a nearby flux calibrator 
star $\alpha$~Lac (HD~213558; A1~V) with single exposure times of 1--2 and 
8--12 sec, respectively.  

The spectra were reduced, extracted, and combined using a 
version of the IDL Spextool software \citep{Cushing2004} modified for 
TripleSpec\footnote{https://www.apo.nmsu.edu/arc35m/Instruments/TRIPLESPEC/TspecTool/index.html}. 
The pipeline includes flat field division, wavelength calibration based upon 
the atmospheric airglow emission lines in the stellar spectra, and spectrum extraction.  
The atmospheric telluric lines were removed and a flux calibration applied using the IDL 
code Xtelluric \citep{Vacca2003}.  This procedure uses a high spectral resolving power model spectrum
of the A0~V star Vega that is fit to the spectrum of the flux calibrator $\alpha$~Lac 
to remove the stellar lines, and then the normalized result is used to extract the 
atmospheric telluric lines.  The final step is to set the absolute flux calibration by 
transforming the model Vega spectrum into a representation of calibrator star spectrum 
by scaling and reddening according to the calibrator star's $B$ and $V$ magnitudes.
However, small differences between the $\alpha$~Lac (A1~V) and Vega (A0~Va) spectra can 
amount to large uncertainties in the estimated flux in the near-infrared part of the spectrum. 
We checked the NIR flux estimates by comparing the transformed Vega spectrum with 
observed fluxes for the calibrator star $\alpha$~Lac from published photometry collected in 
the VizieR Photometry Tool\footnote{https://vizier.cds.unistra.fr/vizier/sed/}
by Anne-Camille Simon and Thomas Boch.  We found that the transformed Vega spectrum
used by Xtelluric to model the spectrum of $\alpha$~Lac actually overestimated the 
observed flux by about $12\%$ in the  $JHK$ bands, so we applied a wavelength-dependent
correction to the RW~Cep fluxes to account for the discrepancy between the applied and 
actual fluxes of the calibrator star $\alpha$~Lac. The final spectrum of RW~Cep is shown in 
Figure~7.  We estimate that the absolute flux calibration has an uncertainty of approximately $10\%$
based upon the scatter between sets of observations and the errors introduced in setting 
fluxes from the calibrator star $\alpha$~Lac.

\begin{figure*}[h]
\plotone{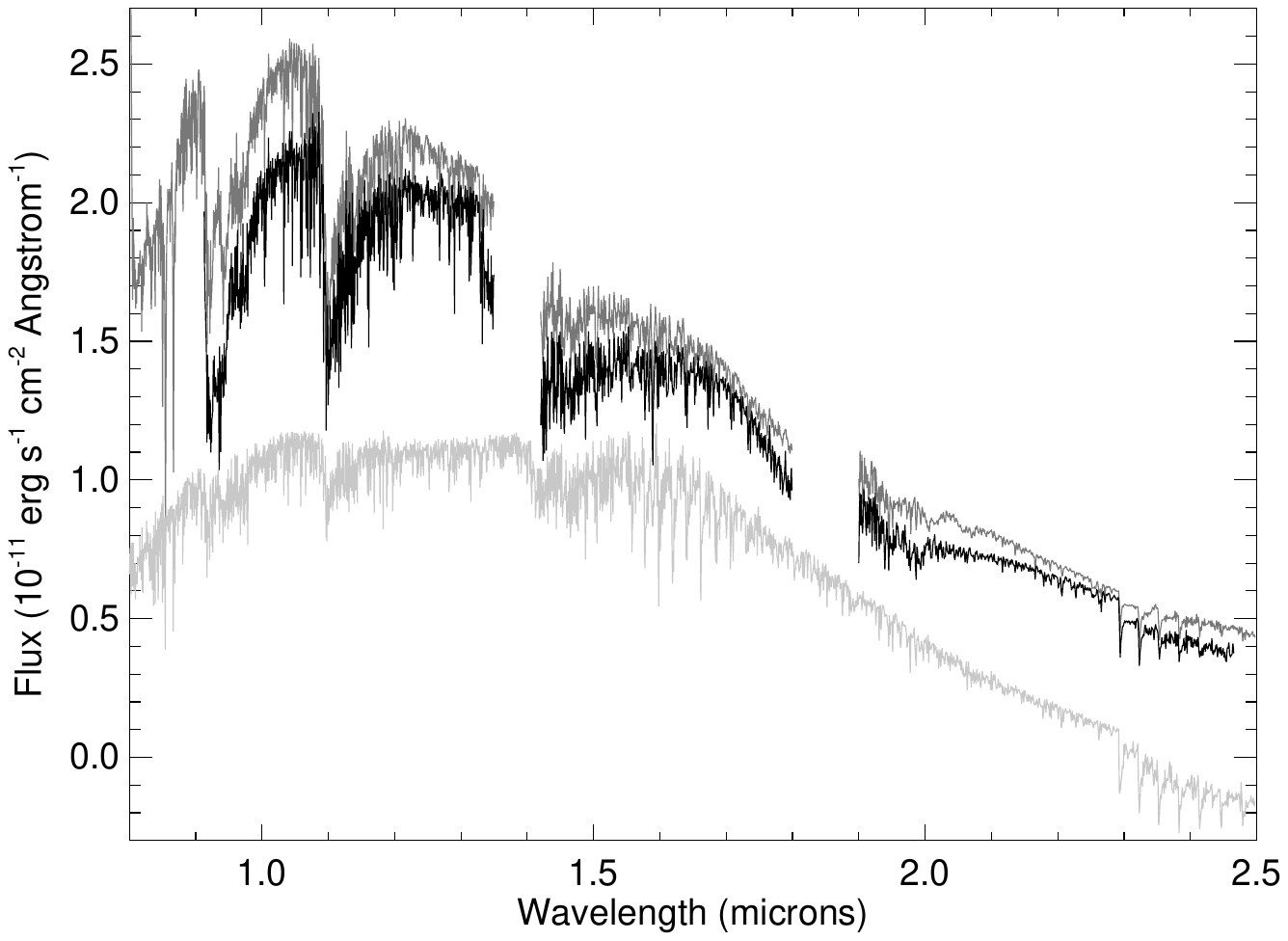}
\caption{The near-IR spectrum of RW~Cep made during the dimming event
(from APO; shown in black) compared to an archival spectrum 
associated with normal brightness (from the IRTF Spectral Library; 
shown in dark gray). A model BT-Dusty spectrum is shown (light gray) 
with an offset of $-0.5 \times 10^{-11}$ erg s$^{-1}$ cm$^{-2}$ \AA $^{-1}$
(see Section 4).  All the spectra are rebinned to a resolving power of $R=1000$ 
for ease of comparison. 
}
\label{fig7}
\end{figure*}

\placefigure{fig7} 

A NIR spectrum of RW~Cep in the pre-dimming state (from 2005 August 26) is 
available from the IRTF Spectral Library\footnote{http://irtfweb.ifa.hawaii.edu/$^\sim$spex/IRTF\_Spectral\_Library/} 
\citep{Rayner2009}, and a corrected version of this spectrum is plotted for comparison in Figure~7. 
The original IRTF spectrum was flux calibrated based upon 2MASS magnitudes that 
unfortunately have large uncertainties ($\pm 0.2$ mag) for such a bright target.  
In the next section, we consider fluxes in the bright state from published photometry
(see Table 2 below).  A comparison of the average fluxes over the $JHK$ bands in the 
IRTF spectrum with the bright state photometric values indicates that the IRTF spectrum is 
approximately $15\%$ fainter than expected.  Consequently, we applied a wavelength-dependent
flux correction to bring the IRTF spectrum into consistency with the photometry, and it is 
the corrected version that is plotted in Figure~7.  This spectrum has associated flux uncertainties 
of about $10\%$ (0.1 mag).   

The recent spectrum made during the dimming event is somewhat fainter than the archival spectrum
by an amount that is larger at shorter wavelength.  The magnitude change from a comparison
of the spectra (Table 2) is $\triangle J = +0.10 \pm 0.19$ mag, $\triangle H = +0.08 \pm 0.19$ mag,
 and $\triangle K = +0.11 \pm 0.25$ mag. 
Together with the visual magnitude estimates (see Figure 1), it appears that 
RW~Cep has faded by approximately 1.1, 0.7, and 0.1 mag in the 
$V$, $I_c$, and $JHK$ bands, respectively (see Figure 9 below).

\begin{figure*}[h]
\plotone{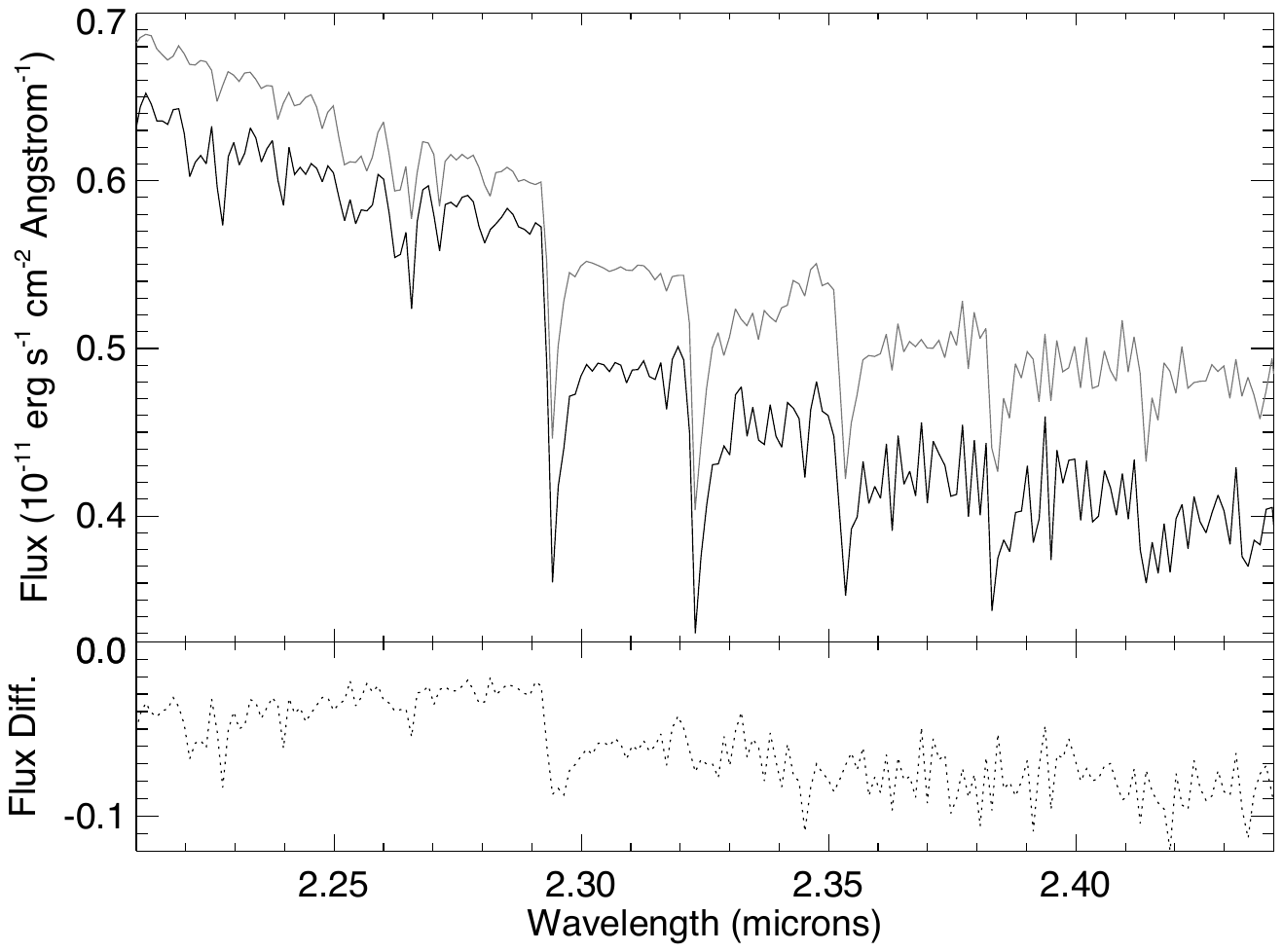}
\caption{The near-IR spectrum of RW~Cep near the CO band-head made during the 
dimming event (from APO; shown in black) compared to an archival spectrum 
associated with normal brightness (from the IRTF Spectral Library; 
shown in dark gray). Both spectra are rebinned to a resolving power of $R=1000$ 
for ease of comparison. The dotted line in the lower panel shows the difference (new minus archival). 
}
\label{fig8}
\end{figure*}

\placefigure{fig8} 

The pre-dimming and dimming event spectra appear similar, but there are 
several significant differences.  We see that the continuum slope in the $K$-band 
is less steep in the dimming event spectrum, and this suggests that there is 
an additional flux component now present that increases in strength with wavelength, 
as expected for dust emission.  
Several of the absorption lines appear somewhat deeper implying a slightly cooler photospheric
temperature.  In particular, the CO 2.29 $\mu$m absorption is now much deeper 
than in the archival spectrum (Figure~8).  \citet{Messineo2021} discuss a number of absorption 
features in the spectra of late-type giants and supergiants (including RW~Cep) that are sensitive 
to stellar effective temperature, and they find that the CO feature grows quickly in 
strength with declining temperature (see their Figure 11, top left panel).  Based 
upon their fit of the temperature dependence of the CO line strength, we estimate 
that the photospheric spectrum indicates a drop from 4200~K (for the archival spectrum) 
to 3900~K during the current faint state (or somewhat less if the absorption strength is 
reduced by dust emission in the $K$-band).

We can use the absolute fluxes from Figure 7 to make approximate estimates
of the temperature distributions associated with the interferometric images. 
The observed flux is related to the angular integral of the image specific intensity:
$$ F_\lambda = \oint I_\lambda  d\omega = \sum I_i \triangle \omega$$
where $I_i$ is the specific intensity of pixel $i$ and $\triangle \omega$ is the angular area 
of each pixel in the image ($\triangle \omega=2.35\times 10^{-19}$ str for $0.1\times0.1$ mas pixels in the SQUEEZE images). 
The observed fluxes $F_\lambda$ averaged over the MIRC-X $H$-band and MYSTIC $K$-band wavelength ranges are
$F_\lambda=(1.36\pm0.17)\times 10^{-11}$ and 
$(6.25\pm1.23)\times 10^{-12}$ erg~sec$^{-1}$~cm$^{-2}$~\AA $^{-1}$, respectively.  
We need to deredden these fluxes to account for interstellar extinction.
Below we derive a faint state reddening of $E(B-V)=0.64 \pm 0.08$ mag (Table 2), 
and this corresponds to NIR extinctions of $A_H=0.34 \pm 0.04$ mag and  
$A_K=0.23 \pm 0.03$ mag \citep{Fitzpatrick1999}. 
Then the extinction-corrected (unreddened) fluxes are $F_\lambda^{UR} = 
(1.86\pm  0.24)\times 10^{-11}$ and $(7.71 \pm 1.51)\times 10^{-12}$ erg~sec$^{-1}$~cm$^{-2}$~\AA $^{-1}$
for the $H$ and $K$-bands, respectively.   We make the simplifying approximation 
that the specific intensities are set by the gas temperature through the Planck function.
We can then use the above equation for the flux to relate the image pixel intensity $P_i$ 
(normalized so that the flux summed over the image is one) to the gas temperature $T$:
$$ T(P_i) = b_2 / \ln (1 + b_1  / P_i).$$  
The constants are $b_1 = {{2hc^2} \over {\lambda ^5}} {{\triangle \omega} \over {F_\lambda^{UR}}} 
= 0.0137 \pm 0.0018$ and $0.0070 \pm 0.0014$ and $b_2 = {{hc} \over {\lambda  k}} = 8909$~K and 6540~K 
for adopted central wavelengths of 1.615 and 2.200 $\mu$m, respectively.
Note the temperatures derived this way may be slight overestimates because part of the 
observed flux may arise in the circumstellar environment and not in the photosphere. 
This method applied to the SQUEEZE images in Figure 5 leads to peak temperatures of around 
4490~K ($H$-band) and 4860~K ($K$-band) averaged over pixels with intensities greater than 
$70\%$ of the maximum intensity.  Similarly, the full disk temperatures are approximately
3520~K for both the $H$ and $K$-bands averaged over pixels with intensities greater than 
$10\%$ of the maximum intensity.  These temperature estimates are similar to that estimated 
above from the CO line (3900~K).  


\section{Spectral Energy Distribution} \label{sec:sed}

We can obtain another estimate of the angular diameter of RW~Cep from 
a comparison of the observed and model flux distributions, but keeping in 
mind that the observed fluxes are actually the sum of stellar and circumstellar 
light.  The shape of the spectral energy distribution (SED) is a 
function of the stellar flux, dust emission, and extinction, so an 
examination of the SED in both the bright and faint states offers a 
means to check on extinction changes resulting from additional circumstellar dust. 
Here we first present the bright state SED based upon archival photometry 
and then compare it to the faint state case based upon current flux estimates.

The fluxes for the bright state were collected from published photometry
collected in the VizieR Photometry Tool.  We added to this set the fluxes
derived from the photometry catalog of \citet{Ducati2002} using the 
flux calibrations from \citet{Bessell1998}. We removed the 2MASS $JHK$ fluxes
that are suspect for this very bright star.  The observed SED is shown in Figure~9 
in the $(\log \lambda, \log \lambda F_\lambda)$ plane.
The measurements indicated by plus signs for wavelengths $< 3$ $\mu$m 
were used in the subsequent fit while the long wavelength
points shown as triangles were omitted because a large fraction of the 
IR excess originates in circumstellar dust \citep{Jones2023}.  We list in column 3 of Table~2 
the averages of the flux measurements in the primary photometric bands. 

\begin{figure*}[h]
    \plotone{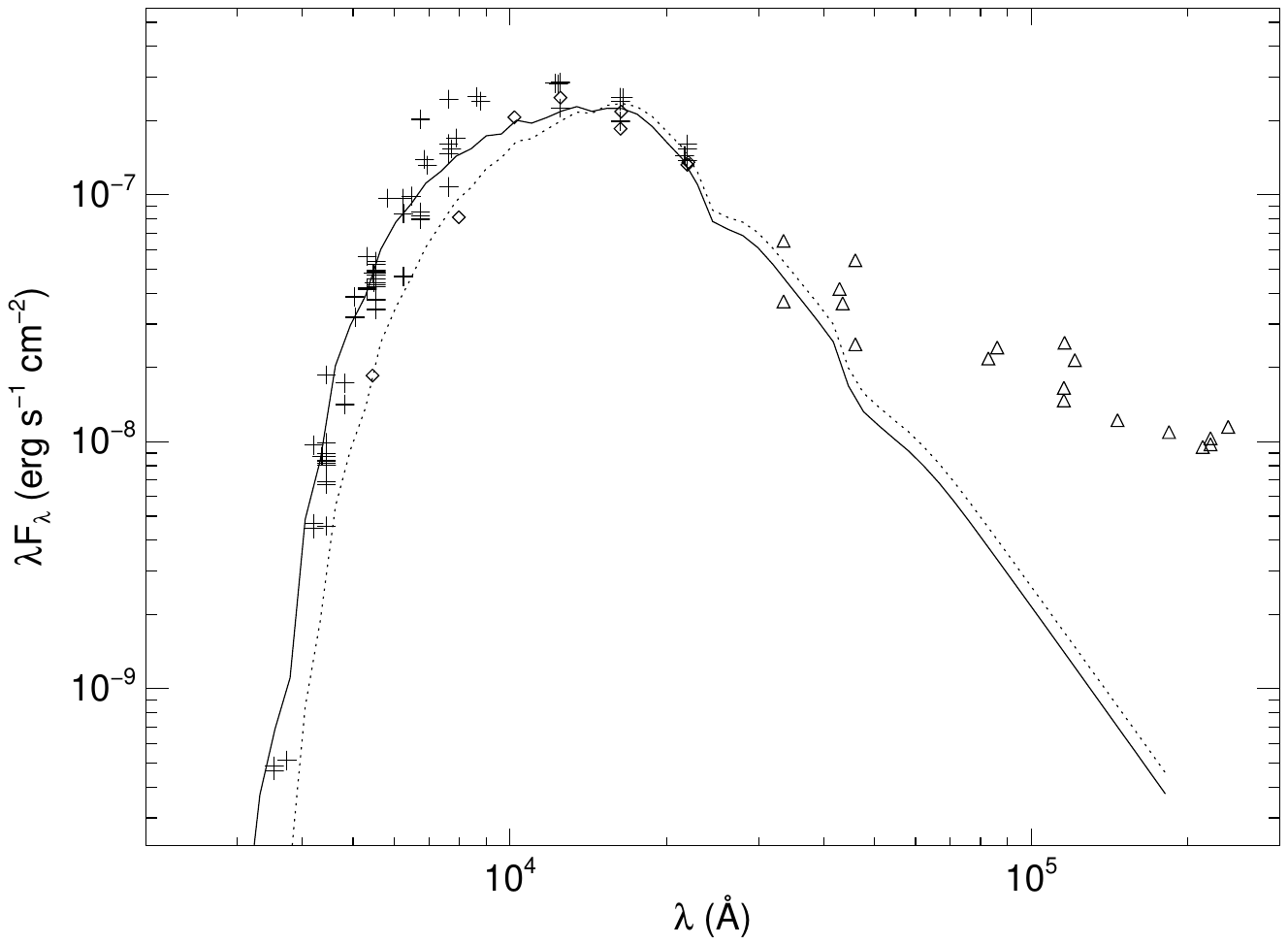}
\caption{The spectral energy distribution of RW~Cep.  The bright state 
fluxes are indicated by plus signs and triangles, and only the former
were used for the fit of the model photospheric spectrum (solid line). 
The diamonds depict the estimated fluxes during the current faint state 
(see Table 2), and the dotted line shows the model photospheric fit for 
the faint state. See Figure~1 from \citet{Jones2023} for the longer wavelength 
part of the SED.
}
\label{fig9}
\end{figure*}

\placefigure{fig9} 

\begin{deluxetable*}{ccccc}[h]
\tablecaption{Spectral Energy Distribution \label{tab2} }
\tablewidth{0pt}
\tablehead{
\colhead{Filter} & 
\colhead{Wavelength}  &
\colhead{Bright State} &
\colhead{Faint State} &
\colhead{Faint State}
\\ 
\colhead{Band}  & 
\colhead{(\AA )}  &   
\colhead{(erg cm$^{-2}$ sec$^{-1}$ \AA $^{-1}$) } &
\colhead{(erg cm$^{-2}$ sec$^{-1}$ \AA $^{-1}$) }  &
\colhead{Source}
}
\startdata
$V$        &  \phn5450 & $(8.16 \pm 0.10)\times 10^{-12}$ & $(3.40 \pm 0.34)\times 10^{-12}$ & AAVSO \\
$I_c$      &  \phn7980 & $(2.07 \pm 0.20)\times 10^{-11}$ & $(1.02 \pm 0.10)\times 10^{-11}$ & KWS \\
$Y$        & 10200 & \nodata                          & $(2.02 \pm 0.25)\times 10^{-11}$  & APO \\
$J$        & 12500 & $(2.17 \pm 0.25)\times 10^{-11}$ & $(1.98 \pm 0.21)\times 10^{-11}$ & APO  \\
$H$        & 16300 & $(1.43 \pm 0.14)\times 10^{-11}$ & $(1.36 \pm 0.17)\times 10^{-11}$ & APO  \\
$H$        & 16300 & \nodata & $(1.14 \pm 0.16)\times 10^{-11}$ & MIRC-X \\
$K$        & 22000 & $(6.82 \pm 0.43)\times 10^{-12}$ & $(6.25 \pm 1.21)\times 10^{-12}$ & APO  \\
$K$        & 22000 & \nodata & $(6.05 \pm 0.96)\times 10^{-12}$ & MYSTIC \\
\hline
$T_{\rm eff}$ (K)  & \nodata & 4200  & 3900 & \nodata \\
$E(B-V)$ (mag)  & \nodata & $0.46 \pm 0.06$                & $0.64 \pm 0.08$ & \nodata \\
$\theta$ (mas)  & \nodata & $2.25 \pm 0.18$                & $2.58 \pm 0.16$ & \nodata \\
\enddata
\end{deluxetable*}

A model of the spectral flux for RW~Cep was selected from the grid of 
BT-Dusty/Phoenix stellar atmosphere models from \citet{Allard2012} 
that are available from the Spanish Virtual Observatory Theoretical Spectra 
Web Server\footnote{http://svo2.cab.inta-csic.es/theory/newov2/}.
These flux distributions are derived from spherical geometry atmospheres 
that use solar abundances from \citet{Asplund2009} and that account for 
mixing processes to create condensates.  We chose a model with 
$T_{\rm eff} = 4200$~K (close to the value of 4185~K derived by  
\citealt{Messineo2021} from spectral indices) and $\log g = 0.5$ (cgs units). 
This gravity value is the lowest in the grid, but it is probably still 
too large by as much as 1 dex for this hypergiant.  However, the 
characteristics of the model spectrum are primarily determined by the 
temperature, so this approximation is reasonable.  The model spectrum
was rebinned to a low resolving power of $R=8$ in order to compare the 
model fluxes with those derived from broad-band photometry. 

The model spectrum was fit to the observed fluxes using two parameters, 
the reddening $E(B-V)$ and the limb-darkened angular diameter $\theta_{LD}$.
We used the reddening law from \citet{Fitzpatrick1999} for a ratio of 
total-to-selective extinction of 3.1.  The fitted model spectrum is 
shown in Figure~9 as a solid line for $E(B-V)=0.46 \pm 0.06$ mag and 
$\theta = 2.25 \pm 0.18$ mas.  \citet{Messineo2021} derive a $K$-band 
extinction of $A(K) = 0.26 \pm 0.17$ mag that corresponds to a reddening
of $E(B-V)=0.72 \pm 0.47$ mag for the reddening law from \citet{Fitzpatrick1999},
which agrees within uncertainties with our result.  The angular diameter 
from the fit is similar to that from the JMMC Stellar Diameters Catalogue
of $\theta_{LD} = 2.14 \pm 0.17$ mas.

Estimates of the flux of RW~Cep in the faint state are listed in 
column 4 of Table~2. The entries for the $V$ and $I_c$ bands are from 
recent AAVSO and KWS magnitudes, respectively, that were converted to 
fluxes using the calibrations from \citet{Bessell1998}.  There are 
four rows that give the average fluxes over the standard filter band
ranges that were obtained from the APO TripleSpec spectrum shown in Figure~7.
In addition, there are estimates for the $H$ and $K$-bands that we derived 
from the raw counts in the CHARA Array observations of RW~Cep and 
the calibrator star HD~219080 (7 And).  A comparison of the detector 
counts from the MIRC-X and MYSTIC observations led to magnitude 
differences of RW~Cep relative to 7~And of $\triangle H = -1.31 \pm 0.14$ mag
and $\triangle K = -1.73 \pm 0.16$ mag. We adopted magnitudes for 7~And of 
$H = 3.81$ and $K=3.77$ from 
\citet{Ducati2002}\footnote{Magnitudes collected from \citet{Gezari1993} which have typical uncertainties of $\pm 0.02$ mag for bright stars.
These uncertainties are much smaller than the other sources of uncertainty in the magnitude error budget.}
so the estimated magnitudes 
of RW~Cep on 2022 December 23 are $H=2.50 \pm 0.14$ mag and $K=2.04 \pm 0.16$ mag.
The corresponding fluxes from the calibrations of \citet{Bessell1998} are 
listed in rows 6 and 8 of Table~2.  

The faint state fluxes are shown as diamond symbols in the SED in Figure~9. 
The flux decrease in the near-IR bands is modest compared to the large drop
observed in the visual $V$-band. We made a separate fit of the faint state 
fluxes this time using a BT-Dusty model for a photospheric temperature of
$T_{\rm eff} = 3900$~K that was indicated by the strength of the CO-band 
features in the TripleSpec observation (Section 3).  This fit is shown as 
the dotted line in Figure~9, and the fitting parameters are 
$E(B-V) = 0.64 \pm 0.08$ mag and $\theta = 2.58 \pm 0.16$ mas. 
Note that these last two parameters are relatively independent of 
the value adopted for the temperature.
For example, if we adopt instead a flux model for $T_{\rm eff} = 4200$~K,
then the fit of the faint state fluxes yields
$E(B-V) = 0.88 \pm 0.07$ mag and $\theta = 2.54 \pm 0.15$ mas.
The angular size of RW~Cep for the faint state flux fits is marginally larger 
than that for the bright state (by $1.4 \sigma$), but we caution that
the fits do not account for the flux component from circumstellar dust. 
The difference in reddening and extinction between the bright and faint 
states is more significant, and it suggests that the current Great Dimming 
of RW~Cep is mainly the result of increased circumstellar dust obscuration
that is particularly important at shorter wavelength.  


\section{Discussion} \label{sec:discussion}

The CHARA Array interferometric observations and the derived images
resolve the photosphere of RW~Cep for the first time. 
The angular diameter estimates from the uniform disk fits are listed in Table~1. 
The MYSTIC visibilities indicate that  
the star is about $27\%$ larger in the longest wavelength channels 
with $\lambda > 2.3$ $\mu$m.  This spectral range corresponds to that 
where the CO transitions are particularly strong (Figure 8),  and we suggest 
that this flux originates at higher levels in the extended atmosphere, 
making the star appear larger at these wavelengths. 
The star appears somewhat box-like in shape in both the $H$ and 
$K$-band SQUEEZE images of Figure~5. 
The range in diameter estimates from the  SQUEEZE images is given in rows 4 and 5 of Table 2. 
The uniform disk diameters from the OITOOLS images are given in rows 6 and 7, and
the limb-darkened diameters associated with SURFING images are  
listed in rows 8 and 9.  The diameters from the OITOOLS and SURFING images 
occupy the mid-range of the estimates found by other methods. 

The distance to RW~Cep is not well established. It is often assumed to be
a member of the Cep~OB1 association \citep{Melnik2020} at a distance of 
3.4~kpc \citep{Rate2020}. It may be a part of the Berkeley~94 star cluster 
at a distance of 3.9~kpc \citep{Delgado2013}.  These estimates agree with the 
distance from Gaia DR2 of $3.4^{+1.4}_{-0.8}$~kpc \citep{BailerJones2018}, but 
are significantly lower than the most recent estimate from Gaia EDR3 of 
$6.7^{+1.6}_{-1.0}$~kpc \citep{BailerJones2021}.  The latter distance would 
place RW~Cep in the Norma/Outer Arm of the Milky Way Galaxy. The discrepancy 
between the Gaia estimates may be related to photocenter jitter related to 
stellar convection and outflows \citep{Chiavassa2022}. We will assume that the 
actual distance falls in the Gaia range of 3.4 to 6.7~kpc.  If we adopt the 
angular diameter from the SURFING images of $\theta_{LD} = 2.45$ mas, then the 
stellar radius is $900 - 1760~R_\odot$ or $4.2 - 8.2$ AU.  
This places RW~Cep among the largest stars known in the Milky Way \citep{Levesque2005}. 

The most striking features in the reconstructed images are the large variations 
in brightness across the visible hemisphere of the star. The surface flux distribution 
is asymmetric with a bright region offset from center and a darker zone towards the western 
side.  The darker zone is slightly more prominent in the $K$-band images,
and the contrast between dark and bright zones may indicate that 
the darker region is related to cool circumstellar dust.  
However, the details of the reconstructed images depend upon assumptions 
about the extended flux, and the images shown in Figures 5 and 6 are representative
of the range in results. 

The NIR spectroscopy shows that the fading is much smaller at longer wavelengths 
compared to that in the visual spectrum. The relative fractions of flux fading from 
the $V$-band (0.55 $\mu$m) to the $K$-band (2.2 $\mu$m) are consistent with 
an extra component of dust extinction with an associated additional reddening of 
$\triangle E(B-V) \approx 0.18$ mag (assuming the nominal extinction law presented by 
\citealt{Fitzpatrick1999}).  Furthermore, the $K$-band continuum slope 
indicates the presence of a dust flux component that contributes progressively 
more flux at longer wavelength.  We suggest that the apparent disk asymmetry 
observed in the interferometric images is also related to this component
of circumstellar dust. 

The Great Dimming of RW~Cep may be the latest in a series of mass ejections
over the last century. \citet{Jones2023} recently presented an analysis of archival 
measurements of the SED of RW~Cep that documents the infrared excess from dust emission.
The SED has one excess component that contributes strongly in the 5 -- 12 $\mu$m range 
and a second component beyond 20 $\mu$m, and \citet{Jones2023} suggest that these 
correspond to inner and outer shells of temperatures 250~K and 100~K, respectively. 
These dust shells have an angular radius of 300 -- 400 mas in an image made at 11.9 $\mu$m
(see their Figure~1). Thus, the current fading may be the latest of continuing mass 
ejection and dust formation episodes, and the newly formed dust now partially 
obscures the visible hemisphere. 

The overall appearance of the $H$ and $K$-band images of RW~Cep 
is similar in character to the asymmetry found by \citet{Montarges2021} 
in visible band images made during the great dimming of Betelgeuse that
they attribute to dust formation in mass ejected from the star.  Furthermore, 
the current dimming of RW~Cep is similar in amplitude and reddening 
to that observed for Betelgeuse (Figure 1).  We suspect that similar processes 
are causing the asymmetric appearance of the CHARA images made during the 
great dimming of RW~Cep.  We note that the star attained a relative brightness 
maximum in 2019 November (JD 2458800 in Figure~1) and then generally faded to its 
current historic minimum \citep{Vollmann2022}.  We suggest that the maximum light 
time may have corresponded to a particularly energetic convective upwelling of hot gas 
that launched a surface mass ejection event.  This gas is now cooling to the 
point of dust formation, and the part of the ejected cloud seen in projection
against the photosphere causes the darker appearance of the western side  
of the  star.  The duration of such dimming events may scale with stellar and dust cloud size, 
so that the timescale ranges from about a year in smaller Betelgeuse, through 
several years for RW~Cep, to decades for larger VY~CMa \citep{Humphreys2021}.
We plan to continue CHARA Array observations over the next year 
to explore how developments in the images are related to the photometric variations.
 
\vspace{0.5 cm}
{\nolinenumbers
This work is based upon observations obtained with the Georgia State University 
Center for High Angular Resolution Astronomy Array at Mount Wilson Observatory.  
The CHARA Array is supported by the National Science Foundation under Grant No.\
AST-1636624, AST-1908026, and AST-2034336.  Institutional support has been provided 
from the GSU College of Arts and Sciences and the GSU Office of the Vice President 
for Research and Economic Development. 
F.B.\ acknowledges funding from the National Science Foundation under Grant No.\ AST-1814777.
S.K.\ acknowledges support from the European Research Council through a Starting Grant 
(Grant Agreement No.\ 639889) and Consolidator Grant (Grant Agreement ID 101003096).
J.D.M.\ acknowledges funding for the development of MIRC-X (NASA-XRP NNX16AD43G, 
NSF AST-1909165) and MYSTIC (NSF ATI-1506540, NSF AST-1909165).
The work is also based on observations obtained with the Apache Point Observatory 3.5-meter 
telescope, which is owned and operated by the Astrophysical Research Consortium.
We thank Russet McMillan and Candace Gray for their help with obtaining the APO observations.  }

\facility{CHARA, APO}
\software{PMOIRED \citep{Merand2022}, SQUEEZE \citep{Baron2010,Baron2012}, SURFING \citep{Roettenbacher2016}, 
Spextool \citep{Cushing2004}, Xtelluric \citep{Vacca2003}} 

\bibliography{ms3.bib}{}

\begin{thebibliography}{}
\expandafter\ifx\csname natexlab\endcsname\relax\def\natexlab#1{#1}\fi
\providecommand{\url}[1]{\href{#1}{#1}}
\providecommand{\dodoi}[1]{doi:~\href{http://doi.org/#1}{\nolinkurl{#1}}}
\providecommand{\doeprint}[1]{\href{http://ascl.net/#1}{\nolinkurl{http://ascl.net/#1}}}
\providecommand{\doarXiv}[1]{\href{https://arxiv.org/abs/#1}{\nolinkurl{https://arxiv.org/abs/#1}}}

\bibitem[{{Allard} {et~al.}(2012){Allard}, {Homeier}, \&
  {Freytag}}]{Allard2012}
{Allard}, F., {Homeier}, D., \& {Freytag}, B. 2012, Philosophical Transactions
  of the Royal Society of London Series A, 370, 2765,
  \dodoi{10.1098/rsta.2011.0269}

\bibitem[{{Anugu} {et~al.}(2020){Anugu}, {Le Bouquin}, {Monnier}, {Kraus},
  {Setterholm}, {Labdon}, {Davies}, {Lanthermann}, {Gardner}, {Ennis},
  {Johnson}, {Ten Brummelaar}, {Schaefer}, \& {Sturmann}}]{Anugu2020}
{Anugu}, N., {Le Bouquin}, J.-B., {Monnier}, J.~D., {et~al.} 2020, \aj, 160,
  158, \dodoi{10.3847/1538-3881/aba957}

\bibitem[{{Asplund} {et~al.}(2009){Asplund}, {Grevesse}, {Sauval}, \&
  {Scott}}]{Asplund2009}
{Asplund}, M., {Grevesse}, N., {Sauval}, A.~J., \& {Scott}, P. 2009, \araa, 47,
  481, \dodoi{10.1146/annurev.astro.46.060407.145222}

\bibitem[{{Bailer-Jones} {et~al.}(2021){Bailer-Jones}, {Rybizki}, {Fouesneau},
  {Demleitner}, \& {Andrae}}]{BailerJones2021}
{Bailer-Jones}, C.~A.~L., {Rybizki}, J., {Fouesneau}, M., {Demleitner}, M., \&
  {Andrae}, R. 2021, \aj, 161, 147, \dodoi{10.3847/1538-3881/abd806}

\bibitem[{{Bailer-Jones} {et~al.}(2018){Bailer-Jones}, {Rybizki}, {Fouesneau},
  {Mantelet}, \& {Andrae}}]{BailerJones2018}
{Bailer-Jones}, C.~A.~L., {Rybizki}, J., {Fouesneau}, M., {Mantelet}, G., \&
  {Andrae}, R. 2018, \aj, 156, 58, \dodoi{10.3847/1538-3881/aacb21}

\bibitem[{{Baron} {et~al.}(2012){Baron}, {Kloppenborg}, \&
  {Monnier}}]{Baron2012}
{Baron}, F., {Kloppenborg}, B., \& {Monnier}, J. 2012, in Society of
  Photo-Optical Instrumentation Engineers (SPIE) Conference Series, Vol. 8445,
  Optical and Infrared Interferometry III, ed. F.~{Delplancke}, J.~K.
  {Rajagopal}, \& F.~{Malbet}, 84451D

\bibitem[{{Baron} {et~al.}(2010){Baron}, {Monnier}, \&
  {Kloppenborg}}]{Baron2010}
{Baron}, F., {Monnier}, J.~D., \& {Kloppenborg}, B. 2010, in Society of
  Photo-Optical Instrumentation Engineers (SPIE) Conference Series, Vol. 7734,
  Optical and Infrared Interferometry II, ed. W.~C. {Danchi}, F.~{Delplancke},
  \& J.~K. {Rajagopal}, 77342I

\bibitem[{{Bessell} {et~al.}(1998){Bessell}, {Castelli}, \&
  {Plez}}]{Bessell1998}
{Bessell}, M.~S., {Castelli}, F., \& {Plez}, B. 1998, \aap, 333, 231

\bibitem[{{Bourg\'{e}s} {et~al.}(2017){Bourg\'{e}s}, {Mella}, {Lafrasse},
  {Duvert}, {Chelli}, {Le Bouquin}, {Delfosse}, \& {Chesneau}}]{Bourges2017}
{Bourg\'{e}s}, L., {Mella}, G., {Lafrasse}, S., {et~al.} 2017, VizieR Online
  Data Catalog, II/346

\bibitem[{{Chiavassa} {et~al.}(2022){Chiavassa}, {Kudritzki}, {Davies},
  {Freytag}, \& {de Mink}}]{Chiavassa2022}
{Chiavassa}, A., {Kudritzki}, R., {Davies}, B., {Freytag}, B., \& {de Mink},
  S.~E. 2022, \aap, 661, L1, \dodoi{10.1051/0004-6361/202243568}

\bibitem[{{Chiavassa} {et~al.}(2009){Chiavassa}, {Plez}, {Josselin}, \&
  {Freytag}}]{Chiavassa2009}
{Chiavassa}, A., {Plez}, B., {Josselin}, E., \& {Freytag}, B. 2009, \aap, 506,
  1351, \dodoi{10.1051/0004-6361/200911780}

\bibitem[{{Cushing} {et~al.}(2004){Cushing}, {Vacca}, \&
  {Rayner}}]{Cushing2004}
{Cushing}, M.~C., {Vacca}, W.~D., \& {Rayner}, J.~T. 2004, \pasp, 116, 362,
  \dodoi{10.1086/382907}

\bibitem[{{Delgado} {et~al.}(2013){Delgado}, {Djupvik}, {Costado}, \&
  {Alfaro}}]{Delgado2013}
{Delgado}, A.~J., {Djupvik}, A.~A., {Costado}, M.~T., \& {Alfaro}, E.~J. 2013,
  \mnras, 435, 429, \dodoi{10.1093/mnras/stt1311}

\bibitem[{{Ducati}(2002)}]{Ducati2002}
{Ducati}, J.~R. 2002, VizieR Online Data Catalog

\bibitem[{{Dupree} {et~al.}(2022){Dupree}, {Strassmeier}, {Calderwood},
  {Granzer}, {Weber}, {Kravchenko}, {Matthews}, {Montarg{\`e}s}, {Tappin}, \&
  {Thompson}}]{Dupree2022}
{Dupree}, A.~K., {Strassmeier}, K.~G., {Calderwood}, T., {et~al.} 2022, \apj,
  936, 18, \dodoi{10.3847/1538-4357/ac7853}

\bibitem[{{Fitzpatrick}(1999)}]{Fitzpatrick1999}
{Fitzpatrick}, E.~L. 1999, \pasp, 111, 63, \dodoi{10.1086/316293}

\bibitem[{{Gehrz} \& {Woolf}(1971)}]{Gehrz1971}
{Gehrz}, R.~D., \& {Woolf}, N.~J. 1971, \apj, 165, 285, \dodoi{10.1086/150897}

\bibitem[{{Gezari} {et~al.}(1993){Gezari}, {Schmitz}, {Pitts}, \&
  {Mead}}]{Gezari1993}
{Gezari}, D.~Y., {Schmitz}, M., {Pitts}, P.~S., \& {Mead}, J.~M. 1993, {Catalog
  of infrared observations} (NASA)

\bibitem[{{H{\"o}fner} \& {Freytag}(2019)}]{Hofner2019}
{H{\"o}fner}, S., \& {Freytag}, B. 2019, \aap, 623, A158,
  \dodoi{10.1051/0004-6361/201834799}

\bibitem[{{Humphreys} {et~al.}(2021){Humphreys}, {Davidson}, {Richards},
  {Ziurys}, {Jones}, \& {Ishibashi}}]{Humphreys2021}
{Humphreys}, R.~M., {Davidson}, K., {Richards}, A.~M.~S., {et~al.} 2021, \aj,
  161, 98, \dodoi{10.3847/1538-3881/abd316}

\bibitem[{{Humphreys} \& {Jones}(2022)}]{Humphreys2022}
{Humphreys}, R.~M., \& {Jones}, T.~J. 2022, \aj, 163, 103,
  \dodoi{10.3847/1538-3881/ac46ff}

\bibitem[{{Jadlovsk{\'y}} {et~al.}(2023){Jadlovsk{\'y}}, {Krti{\v{c}}ka},
  {Paunzen}, \& {{\v{S}}tefl}}]{Jadlovsky2023}
{Jadlovsk{\'y}}, D., {Krti{\v{c}}ka}, J., {Paunzen}, E., \& {{\v{S}}tefl}, V.
  2023, \na, 99, 101962, \dodoi{10.1016/j.newast.2022.101962}

\bibitem[{{Jones} {et~al.}(2023){Jones}, {Shenoy}, \& {Humphreys}}]{Jones2023}
{Jones}, T.~J., {Shenoy}, D., \& {Humphreys}, R. 2023, Research Notes of the
  American Astronomical Society, 7, 92, \dodoi{10.3847/2515-5172/acd37f}

\bibitem[{{Josselin} \& {Plez}(2007)}]{Josselin2007}
{Josselin}, E., \& {Plez}, B. 2007, \aap, 469, 671,
  \dodoi{10.1051/0004-6361:20066353}

\bibitem[{{Keenan} \& {McNeil}(1989)}]{Keenan1989}
{Keenan}, P.~C., \& {McNeil}, R.~C. 1989, \apjs, 71, 245,
  \dodoi{10.1086/191373}

\bibitem[{{Kravchenko} {et~al.}(2021){Kravchenko}, {Jorissen}, {Van Eck},
  {Merle}, {Chiavassa}, {Paladini}, {Freytag}, {Plez}, {Montarg{\`e}s}, \& {Van
  Winckel}}]{Kravchenko2021}
{Kravchenko}, K., {Jorissen}, A., {Van Eck}, S., {et~al.} 2021, \aap, 650, L17,
  \dodoi{10.1051/0004-6361/202039801}

\bibitem[{{Le Besnerais} {et~al.}(2008){Le Besnerais}, {Lacour}, {Mugnier},
  {Thiebaut}, {Perrin}, \& {Meimon}}]{LeBesnerais2008}
{Le Besnerais}, G., {Lacour}, S., {Mugnier}, L.~M., {et~al.} 2008, IEEE Journal
  of Selected Topics in Signal Processing, 2, 767,
  \dodoi{10.1109/JSTSP.2008.2005353}

\bibitem[{{Le Bouquin} {et~al.}(2009){Le Bouquin}, {Lacour}, {Renard},
  {Thi{\'e}baut}, {Merand}, \& {Verhoelst}}]{LeBouquin2009}
{Le Bouquin}, J.-B., {Lacour}, S., {Renard}, S., {et~al.} 2009, \aap, 496, L1,
  \dodoi{10.1051/0004-6361/200811579}

\bibitem[{{Levesque} {et~al.}(2005){Levesque}, {Massey}, {Olsen}, {Plez},
  {Josselin}, {Maeder}, \& {Meynet}}]{Levesque2005}
{Levesque}, E.~M., {Massey}, P., {Olsen}, K.~A.~G., {et~al.} 2005, \apj, 628,
  973, \dodoi{10.1086/430901}

\bibitem[{{L{\'o}pez Ariste} {et~al.}(2023){L{\'o}pez Ariste}, {Wavasseur},
  {Mathias}, {L{\`e}bre}, {Tessore}, \& {Georgiev}}]{LopezAriste2023}
{L{\'o}pez Ariste}, A., {Wavasseur}, M., {Mathias}, P., {et~al.} 2023, \aap,
  670, A62, \dodoi{10.1051/0004-6361/202244285}

\bibitem[{{Maehara}(2014)}]{Maehara2014}
{Maehara}, H. 2014, J.\ Space Science Informatics Japan, 3, 119

\bibitem[{{Martinez} {et~al.}(2021){Martinez}, {Baron}, {Monnier},
  {Roettenbacher}, \& {Parks}}]{Martinez2021}
{Martinez}, A.~O., {Baron}, F.~R., {Monnier}, J.~D., {Roettenbacher}, R.~M., \&
  {Parks}, J.~R. 2021, \apj, 916, 60, \dodoi{10.3847/1538-4357/ac06a5}

\bibitem[{{Massey} {et~al.}(2023){Massey}, {Neugent}, {Ekstr{\"o}m}, {Georgy},
  \& {Meynet}}]{Massey2023}
{Massey}, P., {Neugent}, K.~F., {Ekstr{\"o}m}, S., {Georgy}, C., \& {Meynet},
  G. 2023, \apj, 942, 69, \dodoi{10.3847/1538-4357/aca665}

\bibitem[{{Melnik} \& {Dambis}(2020)}]{Melnik2020}
{Melnik}, A.~M., \& {Dambis}, A.~K. 2020, \mnras, 493, 2339,
  \dodoi{10.1093/mnras/staa454}

\bibitem[{{M{\'e}rand}(2022)}]{Merand2022}
{M{\'e}rand}, A. 2022, in Society of Photo-Optical Instrumentation Engineers
  (SPIE) Conference Series, Vol. 12183, Optical and Infrared Interferometry and
  Imaging VIII, ed. A.~{M{\'e}rand}, S.~{Sallum}, \& J.~{Sanchez-Bermudez},
  121831N

\bibitem[{{Merrill} \& {Wilson}(1956)}]{Merrill1956}
{Merrill}, P.~W., \& {Wilson}, O.~C. 1956, \apj, 123, 392,
  \dodoi{10.1086/146178}

\bibitem[{{Messineo} {et~al.}(2021){Messineo}, {Figer}, {Kudritzki}, {Zhu},
  {Menten}, {Ivanov}, \& {Chen}}]{Messineo2021}
{Messineo}, M., {Figer}, D.~F., {Kudritzki}, R.-P., {et~al.} 2021, \aj, 162,
  187, \dodoi{10.3847/1538-3881/ac116b}

\bibitem[{{Monnier} {et~al.}(2018){Monnier}, {Le Bouquin}, {Anugu}, {Kraus},
  {Setterholm}, {Ennis}, {Lanthermann}, {Jocou}, \& {ten
  Brummelaar}}]{Monnier2018}
{Monnier}, J.~D., {Le Bouquin}, J.-B., {Anugu}, N., {et~al.} 2018, in Society
  of Photo-Optical Instrumentation Engineers (SPIE) Conference Series, Vol.
  10701, Optical and Infrared Interferometry and Imaging VI, ed. M.~J.
  {Creech-Eakman}, P.~G. {Tuthill}, \& A.~{M{\'e}rand}, 1070122

\bibitem[{{Montarg{\`e}s} {et~al.}(2014){Montarg{\`e}s}, {Kervella}, {Perrin},
  {Ohnaka}, {Chiavassa}, {Ridgway}, \& {Lacour}}]{Montarges2014}
{Montarg{\`e}s}, M., {Kervella}, P., {Perrin}, G., {et~al.} 2014, \aap, 572,
  A17, \dodoi{10.1051/0004-6361/201423538}

\bibitem[{{Montarg{\`e}s} {et~al.}(2021){Montarg{\`e}s}, {Cannon}, {Lagadec},
  {de Koter}, {Kervella}, {Sanchez-Bermudez}, {Paladini}, {Cantalloube},
  {Decin}, {Scicluna}, {Kravchenko}, {Dupree}, {Ridgway}, {Wittkowski},
  {Anugu}, {Norris}, {Rau}, {Perrin}, {Chiavassa}, {Kraus}, {Monnier},
  {Millour}, {Le Bouquin}, {Haubois}, {Lopez}, {Stee}, \&
  {Danchi}}]{Montarges2021}
{Montarg{\`e}s}, M., {Cannon}, E., {Lagadec}, E., {et~al.} 2021, \nat, 594,
  365, \dodoi{10.1038/s41586-021-03546-8}

\bibitem[{{Norris} {et~al.}(2021){Norris}, {Baron}, {Monnier}, {Paladini},
  {Anderson}, {Martinez}, {Schaefer}, {Che}, {Chiavassa}, {Connelley},
  {Farrington}, {Gies}, {Kiss}, {Lester}, {Montarg{\`e}s}, {Neilson},
  {Majoinen}, {Pedretti}, {Ridgway}, {Roettenbacher}, {Scott}, {Sturmann},
  {Sturmann}, {Thureau}, {Vargas}, \& {ten Brummelaar}}]{Norris2021}
{Norris}, R.~P., {Baron}, F.~R., {Monnier}, J.~D., {et~al.} 2021, \apj, 919,
  124, \dodoi{10.3847/1538-4357/ac0c7e}

\bibitem[{{Ohnaka} {et~al.}(2011){Ohnaka}, {Weigelt}, {Millour}, {Hofmann},
  {Driebe}, {Schertl}, {Chelli}, {Massi}, {Petrov}, \& {Stee}}]{Ohnaka2011}
{Ohnaka}, K., {Weigelt}, G., {Millour}, F., {et~al.} 2011, \aap, 529, A163,
  \dodoi{10.1051/0004-6361/201016279}

\bibitem[{{Percy} \& {Kolin}(2000)}]{Percy2000}
{Percy}, J.~R., \& {Kolin}, D.~L. 2000, \jaavso, 28, 1

\bibitem[{{Perrin} {et~al.}(2005){Perrin}, {Ridgway}, {Verhoelst}, {Schuller},
  {Coud{\'e} du Foresto}, {Traub}, {Millan-Gabet}, \& {Lacasse}}]{Perrin2005}
{Perrin}, G., {Ridgway}, S.~T., {Verhoelst}, T., {et~al.} 2005, \aap, 436, 317,
  \dodoi{10.1051/0004-6361:20042313}

\bibitem[{{Rate} {et~al.}(2020){Rate}, {Crowther}, \& {Parker}}]{Rate2020}
{Rate}, G., {Crowther}, P.~A., \& {Parker}, R.~J. 2020, \mnras, 495, 1209,
  \dodoi{10.1093/mnras/staa1290}

\bibitem[{{Rayner} {et~al.}(2009){Rayner}, {Cushing}, \& {Vacca}}]{Rayner2009}
{Rayner}, J.~T., {Cushing}, M.~C., \& {Vacca}, W.~D. 2009, \apjs, 185, 289,
  \dodoi{10.1088/0067-0049/185/2/289}

\bibitem[{{Roettenbacher} {et~al.}(2016){Roettenbacher}, {Monnier}, {Korhonen},
  {Aarnio}, {Baron}, {Che}, {Harmon}, {K{\H{o}}v{\'a}ri}, {Kraus}, {Schaefer},
  {Torres}, {Zhao}, {Ten Brummelaar}, {Sturmann}, \&
  {Sturmann}}]{Roettenbacher2016}
{Roettenbacher}, R.~M., {Monnier}, J.~D., {Korhonen}, H., {et~al.} 2016, \nat,
  533, 217, \dodoi{10.1038/nature17444}

\bibitem[{{Rowan-Robinson} \& {Harris}(1982)}]{RowanRobinson1982}
{Rowan-Robinson}, M., \& {Harris}, S. 1982, \mnras, 200, 197,
  \dodoi{10.1093/mnras/200.2.197}

\bibitem[{{Schaefer} {et~al.}(2020){Schaefer}, {ten Brummelaar}, {Gies},
  {Anderson}, {Farrington}, {Golden}, {Jones}, {Klement}, {Majoinen},
  {Ridgway}, {Sturmann}, {Sturmann}, {Turner}, {Vargas}, {Webster}, \&
  {Woods}}]{Schaefer2020}
{Schaefer}, G.~H., {ten Brummelaar}, T.~A., {Gies}, D.~R., {et~al.} 2020, in
  Society of Photo-Optical Instrumentation Engineers (SPIE) Conference Series,
  Vol. 11446, Optical and Infrared Interferometry and Imaging VII, ed. P.~G.
  {Tuthill}, A.~{M{\'e}rand}, \& S.~{Sallum}, 1144605

\bibitem[{{Setterholm} {et~al.}(2022){Setterholm}, {Monnier}, {Le Bouquin},
  {Anugu}, {Ennis}, {Flores}, {Gardner}, {Ibrahim}, {Jocou}, {Kraus},
  {Lanthermann}, {Schaefer}, \& {ten Brummelaar}}]{Setterholm2022}
{Setterholm}, B.~R., {Monnier}, J.~D., {Le Bouquin}, J.-B., {et~al.} 2022, in
  Society of Photo-Optical Instrumentation Engineers (SPIE) Conference Series,
  Vol. 12183, Optical and Infrared Interferometry and Imaging VIII, ed.
  A.~{M{\'e}rand}, S.~{Sallum}, \& J.~{Sanchez-Bermudez}, 121830B

\bibitem[{{Shenoy}(2016)}]{Shenoy2016}
{Shenoy}, D.~P. 2016, PhD thesis, University of Minnesota

\bibitem[{{Taniguchi} {et~al.}(2022){Taniguchi}, {Yamazaki}, \&
  {Uno}}]{Taniguchi2022}
{Taniguchi}, D., {Yamazaki}, K., \& {Uno}, S. 2022, Nature Astronomy, 6, 930,
  \dodoi{10.1038/s41550-022-01680-5}

\bibitem[{{ten Brummelaar} {et~al.}(2005){ten Brummelaar}, {McAlister},
  {Ridgway}, {Bagnuolo}, {Turner}, {Sturmann}, {Sturmann}, {Berger}, {Ogden},
  {Cadman}, {Hartkopf}, {Hopper}, \& {Shure}}]{tenBrummelaar2005}
{ten Brummelaar}, T.~A., {McAlister}, H.~A., {Ridgway}, S.~T., {et~al.} 2005,
  \apj, 628, 453, \dodoi{10.1086/430729}

\bibitem[{{Thi{\'e}baut} \& {Young}(2017)}]{Thiebaut2017}
{Thi{\'e}baut}, {\'E}., \& {Young}, J. 2017, Journal of the Optical Society of
  America A, 34, 904, \dodoi{10.1364/JOSAA.34.000904}

\bibitem[{{Tsuji}(2006)}]{Tsuji2006}
{Tsuji}, T. 2006, \apj, 645, 1448, \dodoi{10.1086/504585}

\bibitem[{{Vacca} {et~al.}(2003){Vacca}, {Cushing}, \& {Rayner}}]{Vacca2003}
{Vacca}, W.~D., {Cushing}, M.~C., \& {Rayner}, J.~T. 2003, \pasp, 115, 389,
  \dodoi{10.1086/346193}

\bibitem[{{Vollmann} \& {Sigismondi}(2022)}]{Vollmann2022}
{Vollmann}, W., \& {Sigismondi}, C. 2022, The Astronomer's Telegram, 15800, 1

\bibitem[{{Wilson} {et~al.}(2004){Wilson}, {Henderson}, {Herter}, {Matthews},
  {Skrutskie}, {Adams}, {Moon}, {Smith}, {Gautier}, {Ressler}, {Soifer}, {Lin},
  {Howard}, {LaMarr}, {Stolberg}, \& {Zink}}]{Wilson2004}
{Wilson}, J.~C., {Henderson}, C.~P., {Herter}, T.~L., {et~al.} 2004, in Society
  of Photo-Optical Instrumentation Engineers (SPIE) Conference Series, Vol.
  5492, Ground-based Instrumentation for Astronomy, ed. A.~F.~M. {Moorwood} \&
  M.~{Iye}, 1295--1305

\end{thebibliography}
\bibliographystyle{aasjournal}

\end{document}